\documentclass[a4paper,fleqn,usenatbib,useAMS]{mnras}

\usepackage[T1]{fontenc}
\usepackage{ae,aecompl}
\usepackage{natbib}

\usepackage{CJK}
\usepackage{graphicx}	
\usepackage{amsmath}	
\usepackage{amssymb}	
\usepackage{multicol}        
\usepackage{bm}		
\usepackage{pdflscape}	
\usepackage{booktabs}
\usepackage{caption}
\usepackage{mathtools}

\newcommand{\angstrom}{\textup{\AA}}

\newcommand{\fout}{\ensuremath{\langle f_{900}/f_{1500}\rangle_{\rm out}}}
\newcommand{\ziii}{\ensuremath{z\sim3}}
\newcommand{\wlya}{$W_{\lambda}$(Ly$\rm \alpha$)}
\newcommand{\luv}{\ensuremath{L_{\rm UV}}}
\newcommand{\fesca}{\ensuremath{f_{\rm esc,abs}}}
\newcommand{\zspec}{\ensuremath{z_{\rm sys}}}

\title[Uncontaminated LyC]{An uncontaminated measurement of the escaping Lyman continuum at {\ziii}}

\author[Pahl et al.]{Anthony J. Pahl,$^{1}$\thanks{Contact e-mail: \href{mailto:pahl@astro.ucla.edu}{pahl@astro.ucla.edu}}
Alice Shapley,$^{1}$
Charles C. Steidel,$^{2}$
Yuguang Chen (陈昱光),$^{2}$\newauthor
and Naveen A. Reddy$^{3}$
	\\
	$^{1}$Department of Physics and Astronomy, University of California, Los Angeles, CA 90095, USA\\
	$^{2}$Cahill Center for Astronomy and Astrophysics, California Institute of Technology, MC249-17, Pasadena, CA 91125, USA\\
	$^{3}$Department of Physics and Astronomy, University of California Riverside, Riverside, CA 92521, USA\\
}

\date{}

\pubyear{}

\begin{document}
\begin{CJK*}{UTF8}{gbsn}
		
	\label{firstpage}
	\pagerange{\pageref{firstpage}--\pageref{lastpage}}
	\maketitle
	
\begin{abstract}
	Observations of reionization-era analogs at {\ziii} are a powerful tool for constraining reionization. 
	Rest-ultraviolet observations are particularly useful, in which both direct and indirect tracers of ionizing-photon production and escape can be observed.
	We analyse a sample of 124 $z\sim3$ galaxies from the Keck Lyman Continuum Spectroscopic Survey, with sensitive spectroscopic measurements of the Lyman continuum region.
	We present a method of removing foreground contamination from our sample using high-resolution, multi-band \textit{Hubble Space Telescope} imaging. 
	We re-measure the global properties of the cleaned sample of 13 individually-detected Lyman continuum sources and 107 individually-undetected sources, including a sample-averaged absolute escape fraction of $\fesca=0.06\pm0.01$ and a sample-averaged ratio of ionizing to non-ionizing ultraviolet flux density of $\fout=0.040\pm0.006$, corrected for attenuation from the intergalactic and circumgalactic media.
	Based on composite spectra, we also recover a strong positive correlation between {\fout} and Ly$\alpha$ equivalent width ({\wlya}) and a negative correlation between {\fout} and UV luminosity.
	As in previous work, we interpret the relationship between {\fout} and {\wlya} in terms of the modulation of the escape of ionizing radiation from star-forming galaxies based on the covering fraction of neutral gas.
	We also use a {\wlya}-weighted {\fout} to estimate an ionizing emissivity from star-forming galaxies at {\ziii} as $\epsilon_{\rm LyC} \simeq 5.5 \times 10^{24}$~erg~s$^{-1}$~Hz$^{-1}$~Mpc$^{-3}$.
	This estimate, evaluated using the uncontaminated sample of this work, affirms that the contribution of galaxies to the ionizing background at \ziii{} is comparable to that of active galactic nuclei.
\end{abstract}


\defcitealias{Steidel2018}{S18}
\defcitealias{Mostardi2015a}{M15}
\defcitealias{Vanzella2012}{Vanzella et al. 2012}

\section{Introduction} \label{sec:intro}

Reionization, the last major phase transition of the Universe, describes the ionization of the Hydrogen gas in the inter-galactic medium (IGM) in the first billion years of the Universe's history. Since the reionization process ends by $z\sim6$ \citep{Fan2006,Robertson2015a, Planck2016} and the number density of luminous QSOs drops off sharply beyond $z\sim2$ \citep{Shen2020}, the prevailing notion is that ionizing radiation leaking from early star-forming galaxies drives the process of reionization \citep[e.g., ][]{Bouwens2015a, Parsa2018a}. 

A number of observational constraints and challenges prevent a straightforward analysis of the galaxies that contribute strongly to the ionizing background during reionization. While we can chart the abundance of star-formation rates (SFRs) as a function of redshift \citep{Madau2014a}, the difficulty lies in converting the cosmic SFR density ($\rho_{\rm SFR}$) into an ionizing emissivity, a key factor in dictating the evolution of the IGM neutral fraction \citep{Bouwens2015a,Robertson2015a}. 
One of the essential parameters for converting between $\rho_{\rm SFR}$ and the ionizing emissivity is {\fesca}, the absolute escape fraction of ionizing radiation produced in H~\textsc{ii} regions that evades absorption by neutral-phase gas and dust in the ISM. Unfortunately, it is not possible to constrain {\fesca} directly based on measurements during the epoch of reionization considering the drastic drop-off of transmission of ionizing rest-UV continuum through the IGM past $z>3.5$ \citep{Vanzella2012,Steidel2018}. Instead, there has been a focus on sources analogous to the $z>6$ reionizing population but observed at redshifts where the IGM is still transparent to Lyman Continuum (LyC) radiation at $\lambda\leq912$~\AA{}.

Many of the direct {\fesca} measurements in the local Universe come from compact galaxies with large [O~\textsc{iii}]/[O~\textsc{ii}] ratios and high [O~\textsc{iii}]$\lambda5007$ equivalent widths, referred to as ``Green Peas" \citep[HST/COS, ][]{Borthakur2014, Izotov2016a,Izotov2018,Izotov2021}. At higher redshifts ({\ziii}), the properties of galaxies and their circumgalactic environments begin to align more with those in the reionization era. Meanwhile, the IGM opacity is still at a level that allows for the direct detection of LyC emission from star-forming galaxies. These aspects make this redshift range compelling for LyC surveys, but the sightline-to-sightline variability in IGM opacity makes the interpretations of individual LyC detections difficult. Thanks to advances in deep optical spectroscopy and narrowband imaging, there have been a handful of individual LyC detections at {\ziii} \citep[e.g., ][]{Mostardi2015a,Shapley2016a,DeBarros2016,Bian2017,Vanzella2016a,Vanzella2017a}. In order to understand the LyC of the full population of galaxies at {\ziii}, however, larger and representative ensembles of galaxies with sensitive LyC measurements are required, which can smooth out the deviations in the transmission of the IGM through sample averaging.

To this end, the Keck Lyman Continuum Spectroscopic (KLCS) survey \citep[][hereafter S18]{Steidel2018} utilized the LRIS double spectrograph on the Keck I telescope \citep{Oke1995,Steidel2004} to observe 124 $2.75\leq z\leq3.4$ galaxies in the rest-UV with the goal of constructing spectral composites that represent global properties of {\ziii} galaxies. \citetalias{Steidel2018} reported a sample-averaged absolute escape fraction of $\fesca=0.09\pm0.01$ for their sample after performing spectroscopic contamination rejection as well as fitting Binary Model and Stellar Synthesis \citep[BPASS; ][]{Eldridge2017} templates and physically-motivated ISM models to rest-frame UV composite spectra. This sample-averaged {\fesca} has a robust correction for IGM and CGM absorption due to the large sample number that went into the calculation. These results are complemented by other large-scale LyC surveys at $z\sim3-4$, such the LACES survey of 61 $z\sim3.1$ Ly$\rm \alpha$-emitting galaxies using WFC3/UVIS F336W imaging on $HST$ \citep{Fletcher2019} and the survey of 201 star-forming $z\sim4$ galaxies drawn from the VUDS survey using VLT/VIMOS spectroscopy \citep{Marchi2018}.

One remaining limitation in interpreting the results of high-redshift LyC surveys is the risk of non-ionizing UV light from foreground objects masquerading as an apparent ionizing signal. From the ground, a $z\sim3$ galaxy may appear as a coherent object with strong LyC flux density, but higher-resolution imaging can reveal a lower-redshift source nearby along the line of sight that contributes to the observed-frame $\sim$3600~\AA{} spectrum. 
\citet{Vanzella2012} demonstrated the potential pitfalls of line-of-sight contamination in seeing-limited ground-based LyC observations. Building on these ideas, \citet[][hereafter M15]{Mostardi2015a} examined 16 apparent $z\sim2.85$ LyC leakers identified through ground-based narrowband imaging. Multi-band \textit{Hubble Space Telescope} ($HST$) imaging was used to examine the photometric-redshifts of individual sub-components within the seeing-limited extent of each target. \citetalias{Mostardi2015a} found that 15 out of 16 of the apparent LyC detections were contaminated with foreground signal. The demonstrated efficacy of this methodology and the propensity for line-of-sight contamination highlights the importance of high-resolution, space-based follow up of LyC detection candidates at {\ziii}.

In the current work, we attempt to increase the confidence of the results of one of the largest {\ziii} LyC surveys, the KLCS survey, by examining the strongest LyC sources with \textit{HST} imaging. We specifically targeted the 15 KLCS galaxies with significant (3$\sigma$) individual detections of LyC flux in deep Keck/LRIS spectroscopy, as they significantly affect the global statistics of the sample. We simultaneously examined 24 additional galaxies in the sample that fall on the \textit{HST} mosaics but lack individual LyC detections. In addition to searching for foreground contamination in the KLCS sample, this work also serves as an update to \citetalias{Steidel2018} by revisiting the key results of the paper using a cleaned sample. Accordingly, our analysis yields a statistically robust, uncontaminated description of the LyC properties of {\ziii} star-forming galaxies.

In Section \ref{sec:sample_method}, we describe the KLCS parent sample, the ground-based observation methodology, and the details of the \textit{HST} observations. In Section \ref{sec:reduction}, we present the reduction procedures, photometric analysis including the segmentation of objects with complex morphologies, and subsequent photometric measurements. In Section \ref{sec:contam}, we detail the search for foreground contamination and present the results of the contamination analysis. Finally, in Section \ref{sec:redo}, we report updated measurements of the LyC properties of the KLCS sample after contamination removal. We summarize our results in Section \ref{sec:summary}.

Throughout this paper, we adopt a standard $\Lambda$CDM cosmology with $\Omega_m$ = 0.3, $\Omega_{\Lambda}$ = 0.7 and $H_0$ = 70 $\textrm{km\,s}^{-1}\textrm{Mpc}^{-1}$. We also employ the AB magnitude system.

\section{Sample and Methodology} \label{sec:sample_method}

The KLCS survey \citepalias{Steidel2018} utilized the Keck/LRIS spectrograph \citep{Oke1995,Steidel2004} to obtain deep rest-UV spectra of an initial target sample of 137 galaxies selected as Lyman Break Galaxies (LBGs) at $2.75 \leq z \leq 3.4$ \citep[for a description of $U_{\rm n}GR$ selection methodology, see][]{Steidel2003,Adelberger2004,Reddy2012}. 
The LRIS spectra of KLCS galaxies cover the Lyman continuum region (LyC, 880-910~\AA), the Ly$\alpha$ feature, and far-UV interstellar metal absorption features redward of Ly$\alpha$.
The LRIS observations were taken between 2006 and 2008 across nine different survey fields with integration times per mask of $\sim10$ hr.
Of the 137 galaxies observed, eight objects were removed due to instrumental defects or clear evidence for blending with nearby sources. The remaining 2D and 1D spectra were examined for evidence of spectral blending, where an additional redshift along the line of sight could indicate contamination in the apparent LyC flux of a higher-redshift source. Five galaxies were removed due to apparent spectral blending, resulting in a final sample of 124 galaxies for analysis. 
\citetalias{Steidel2018} examined the ratio of ionizing to non-ionizing UV flux density within the KLCS sample, defined as the average flux density within $880 \leq \lambda_{0}/\angstrom \leq 910$ ($f_{900}$) divided by the average flux density within $1475 \leq \lambda_{0}/\angstrom \leq 1525$ ($f_{1500}$), or $\langle f_{900}/f_{1500}\rangle_{\rm obs}$. Fifteen galaxies were significantly detected ($f_{900} > 3\sigma_{900}$, where $\sigma_{900}$ is the $f_{900}$ measurement uncertainty) and were defined as the LyC detection sample. The remaining 109 galaxies with $f_{900} < 3\sigma_{900}$ were defined as the LyC non-detection sample.

\begin{figure} 
	\centering
	\includegraphics[width=\columnwidth]{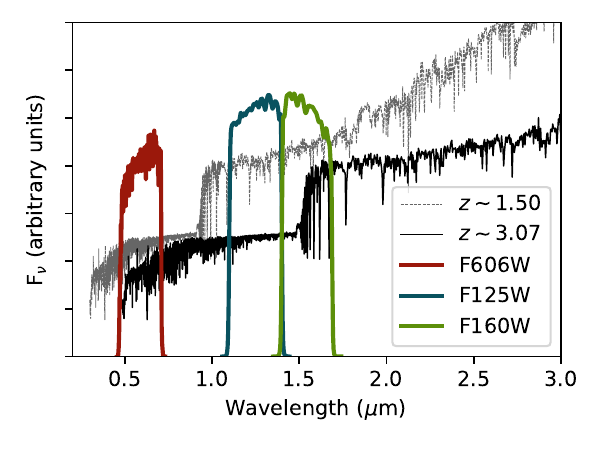}
	\caption{$HST\: V_{606}J_{125}H_{160}$ filter set in wavelength space in relation to a characteristic SED from BPASS \citep{Stanway2018} shown at different redshifts. This SED was generated with 100Myr constant star-formation history and solar metallicity, and was reddened using $E(B-V)=0.2$ and assuming a \citet{Calzetti1999} extinction curve. The black curve indicates the SED shifted to $z=3.07$, the median redshift of the LyC detection sample. At this redshift, the $J_{125}-H_{160}$ color probes the 4000~\AA{} Balmer break. 
	The grey curve shows the SED shifted to $z=1.5$, a redshift typical of low-redshift interlopers. At lower redshift, the $V_{606}-J_{125}$ probes the Balmer break.
	}
	\label{fig:filters}
\end{figure}

Despite significant efforts to remove foreground contamination from the KLCS sample through examination of the 1D and 2D spectra, high-resolution imaging has previously been required to vet individual detections of leaking LyC \citepalias{Vanzella2012,Mostardi2015a}. In {\ziii} LyC detections, apparent LyC leakage at $\sim3600$~\AA{} may actually originate from a  low-redshift component that lies in projection within the angular extent of a clumpy galaxy morphology \citep[e.g., ][]{Vanzella2016a,Siana2015}. Thus, we require high-resolution, space-based imaging for spatially-resolved photometric-redshift analysis of KLCS LyC detections. These observations enable the consideration of individual sub-components as potential contaminants. We require imaging in multiple filters to judge the redshift of individual, extracted components. The $HST\: V_{606}J_{125}H_{160}$ filter set is ideal for this type of analysis \citepalias{Mostardi2015a}. At $z=3.07$, the median redshift of the KLCS detection sample, the $J_{125}$ and $H_{160}$ band filters are situated on either side of the Balmer break. 
These filters are shown in Figure \ref{fig:filters} along with the SED typical of {\ziii} LBGs from the BPASS set of stellar-population models \citep{Stanway2018} redshifted to $z=3.07$. Also displayed is the same BPASS SED at $z=1.50$, representing a low-redshift interloper. At {\ziii}, the $J_{125}-H_{160}$ color is expected to reflect the Balmer break, while at lower redshift, $J_{125}-H_{160}$ color is expected to be flatter as both bands probe redward of the break. Additionally, at {\ziii}, $V_{606}-J_{125}$ is sensitive to the rest-frame UV spectral slope, which provides information on the stellar age and reddening of the galaxy. At lower redshift, the Balmer break passes between the $V_{606}$ and $J_{125}$ bands. 

\begin{figure*} 
	\centering
	\includegraphics[width=\textwidth]{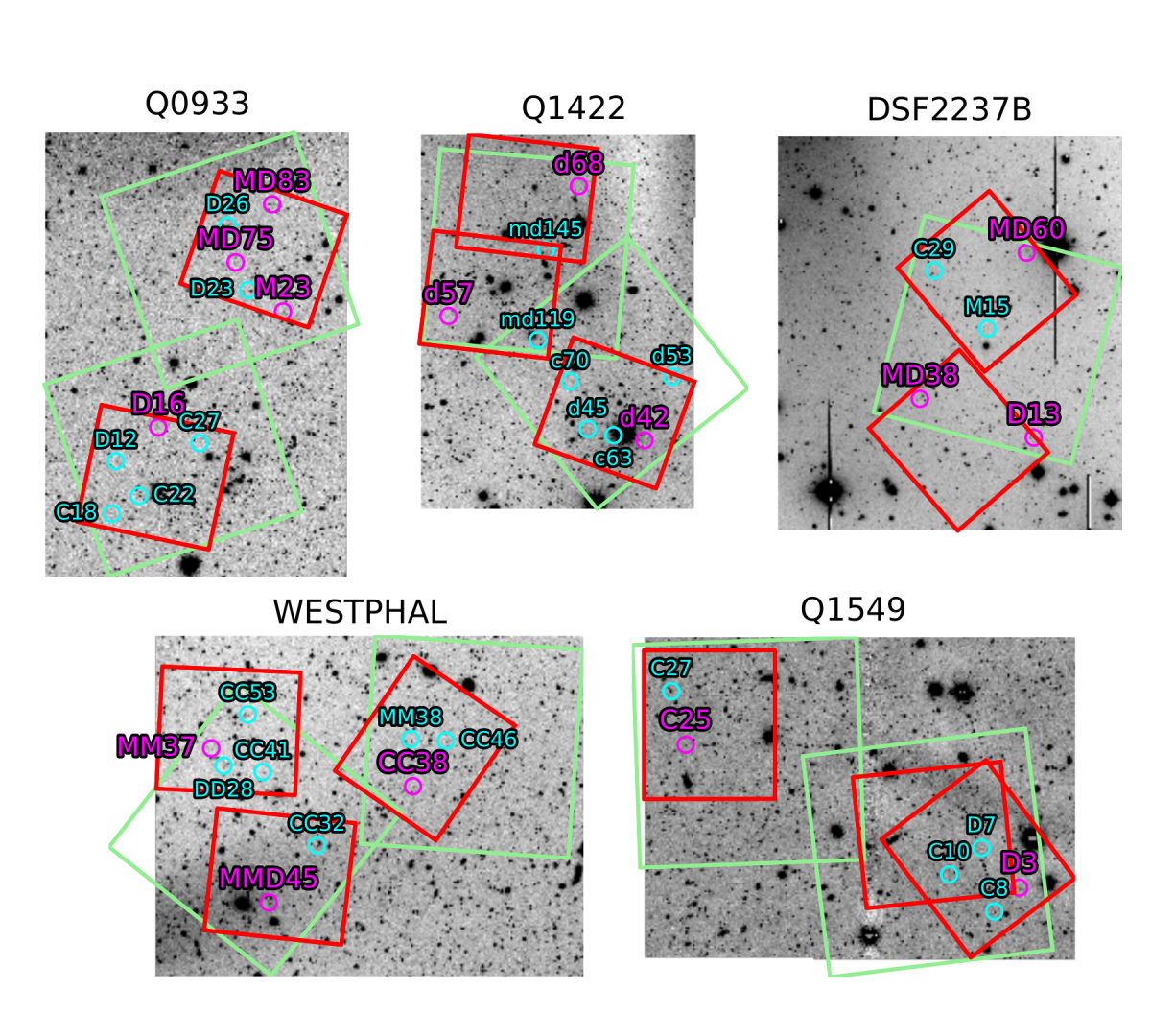}
	\caption{The five fields targeted by our HST observing program. In green are the $202\arcsec\times202\arcsec$ footprints of the ACS pointings, with imaging taken in the $V_{606}$ band. In red are the $136\arcsec\times123\arcsec$ footprints of the WFC3-IR pointings, with imaging taken in $J_{125}$ and $H_{160}$ bands. The magenta circles represent the positions of the 14 LyC detection candidates, while the cyan circles represent the non-detections that lie in the same fields. One LyC detection candidate (Q1549-C25) was not the target of this observing program; the $HST\: V_{606}J_{125}H_{160}$ images are described in \citetalias{Mostardi2015a} \citep[also see][]{Shapley2016a}.
	}
	\label{fig:fields}
\end{figure*}

We observed 14 KLCS LyC detection candidates with ACS/F606W ($V_{606}$), WFC3/F125W ($J_{125}$), and WFC3/F160W ($H_{160}$) on \textit{HST} as a part of Cycle 25 Program ID 15287 (PI: Shapley). The observations took place between 2017 and 2019 in five separate LBG survey fields (Q0933, Q1422, DSF2237b, Q1549, and Westphal) across seven ACS pointings and 11 WFC3 pointings. The observations within each field were designed to cover the 14 LyC detection candidates with apparent LyC detections but no multi-band HST data, whereas Q1549-C25 had already been analysed by \citet{Shapley2016a}. We found the most efficient combination of pointings in each field that covered all apparent LyC detections with $V_{606}$, $J_{125}$, and $H_{160}$. Coincidentally, we covered with all three filters an additional 24 KLCS sources lacking individual LyC detections. These pointings are shown in Figure \ref{fig:fields} alongside KLCS sources covered by all three filters. An additional 11 KLCS LyC non-detections were covered by at least one filter. Each pointing was observed for three orbits per filter, or $\sim7000\:\rm s$ of exposure time. 
The total duration of the program was 87 orbits, required to ensure >10$\sigma$ detections in $V_{606}$ for potential contaminant components, based on the faintest $f_{900}$ (i.e., observed 3600~\AA{} flux density) values of the detection sample and assuming (at worst) a flat SED between $\sim3600$~\AA{} and $V_{606}$. 
We executed each orbit using 4-point DITHER-BOX dither patterns for ACS/WFC pointings and 4-point DITHER-LINE patterns for WFC3/IR pointings.
For one LyC detection, Q1549-D3, we included previously-acquired $HST\; V_{606}J_{125}H_{160}$ imaging from Cycle 20 Program ID 12959 \citepalias[PI Shapley, ][]{Mostardi2015a}. We included 5 orbits of $V_{606}$ in a single pointing and 3 orbits of $J_{125}$ and $H_{160}$ each in a single pointing to cover this detection. We re-reduced these exposures with the same techniques applied to the more recent Cycle 25 data to provide consistency across the datasets.

\section{Reduction} \label{sec:reduction}

In this section, we detail the steps necessary to convert raw \textit{HST} imaging to aligned images and photometric measurements of our targets. Many of the procedures in this section follow those detailed in \citetalias{Mostardi2015a}.

\subsection{Mosaic generation and PSF matching}

First, the calibrated, flux-fielded, and charge-transfer-efficiency-corrected (for ACS) exposures were combined and aligned to produce a single mosaic for each filter and field observed. This reduction step was completed using the \textit{DrizzlePac} software package \citep{Fruchter2010}. Contained in the package is the task \textit{AstroDrizzle}, which performs automatic cosmic-ray rejection and sky subtraction during image combination, as well as the alignment task \textit{TweakReg}.
We used \textit{TweakReg} to calculate residual shifts in RA and Dec. by inputting coordinate lists of extended sources that were first measured using \textit{SExtractor} \citep{Bertin1996}. After calculating the shifts by comparing the coordinate lists, \textit{TweakReg} performs corrections by re-registering the WCS of each exposure.
The individual exposures within each pointing and filter did not require significant re-alignment to one another, save for the F160W imaging of the northern WFC3/IR pointing in the DSF2237b field, in which HST experienced a reaction-wheel failure partway through the exposures and required a $\sim$0.05$\arcsec$ adjustment.

After averaged mosaics were generated for each pointing and filter combination, the mosaics within each of the five target fields were aligned to one another and registered on the same WCS. 
The alignment RMS in both RA and Dec. between ACS images of different pointings was $\sim$0.005$\arcsec$ and the alignment RMS between WFC3/IR and ACS pointings was $\sim$0.01$\arcsec$. The final $V_{606}$, $J_{125}$, and $H_{160}$ mosaics produced by \textit{AstroDrizzle} were generated with matching pixel grids across filters at a pixel scale of 0.03$\arcsec$/px and a drop-to-pixel ratio (\verb|final_pixfrac|) of 0.7 to maximize pixel resolution of the morphological features in the $V_{606}$ imaging. These final mosaics were registered to the WCS of the ground-based $UGR$ images \citep[decribed in][]{Reddy2012} available for these fields using the task \textit{ccmap} in \textsc{IRAF} and manually-generated starlists of $\sim$50 unsaturated stars. The typical RMS in alignment between the \textit{HST} mosaics and the ground-based $G$ imaging in both RA and Dec. was $\sim$0.06$\arcsec$. The LyC detection candidates in both ground-based imaging and aligned \textit{HST} imaging are shown in the Figure \ref{fig:dets}.

For one object, Westphal-CC38, a small area of decreased sensitivity, or ``blob" \citep{Sunnquist2018}, was in the vicinity of the object in the WFC3/IR images. To accurately measure the photometry of this object, we used the sky flats of \citet{Mack2021} that included corrections to the blobs present on the IR detector. 

Next, the $V_{606}$ and $J_{125}$ mosaics were smoothed to the same angular resolution as the corresponding $H_{160}$ mosaic to allow for accurate measurement of colors. Hereafter, we refer to the smoothed $V_{606}$ and $J_{125}$ mosaics as $sm(V_{606})$ and $sm(J_{125})$.
The point-spread functions (PSFs) for each field and filter were generated using a set of 10-15 bright, unsaturated point sources identified manually within each mosaic. We combined these stellar profiles into PSFs using the $Astropy$ routine $Photutils$ \citep{Bradley2020}. The average Gaussian full width at half maximum (GFWHM) of the $V_{606}$, $J_{125}$, and $H_{160}$ PSFs across all fields were 0.100$\arcsec$, 0.177$\arcsec$, and 0.183$\arcsec$, respectively. 
We used the \textsc{IRAF} routine \textit{psfmatch} to generate and apply kernels matching the $V_{606}$ and $J_{125}$ mosaics to the resolution of the $H_{160}$ PSF.
Curve of growth analysis of the $sm(V_{606})$ and $sm(J_{125})$ PSFs with the original $H_{160}$ PSF demonstrated that the fraction of enclosed flux for all three agrees to within 1\% at 0.5$\arcsec$.
We display the PSF-matched $sm(V_{606})$ and $sm(J_{125})$ and original-resolution $H_{160}$ postage stamps of the LyC detection candidates as well as $V_{606}J_{125}H_{160}$ (for blue, green, and red) false-color images in Figure \ref{fig:dets}. Postage stamps of the non-detections can be found in the Appendix.

\begin{figure*} 
	\centering
	\includegraphics[width=\textwidth]{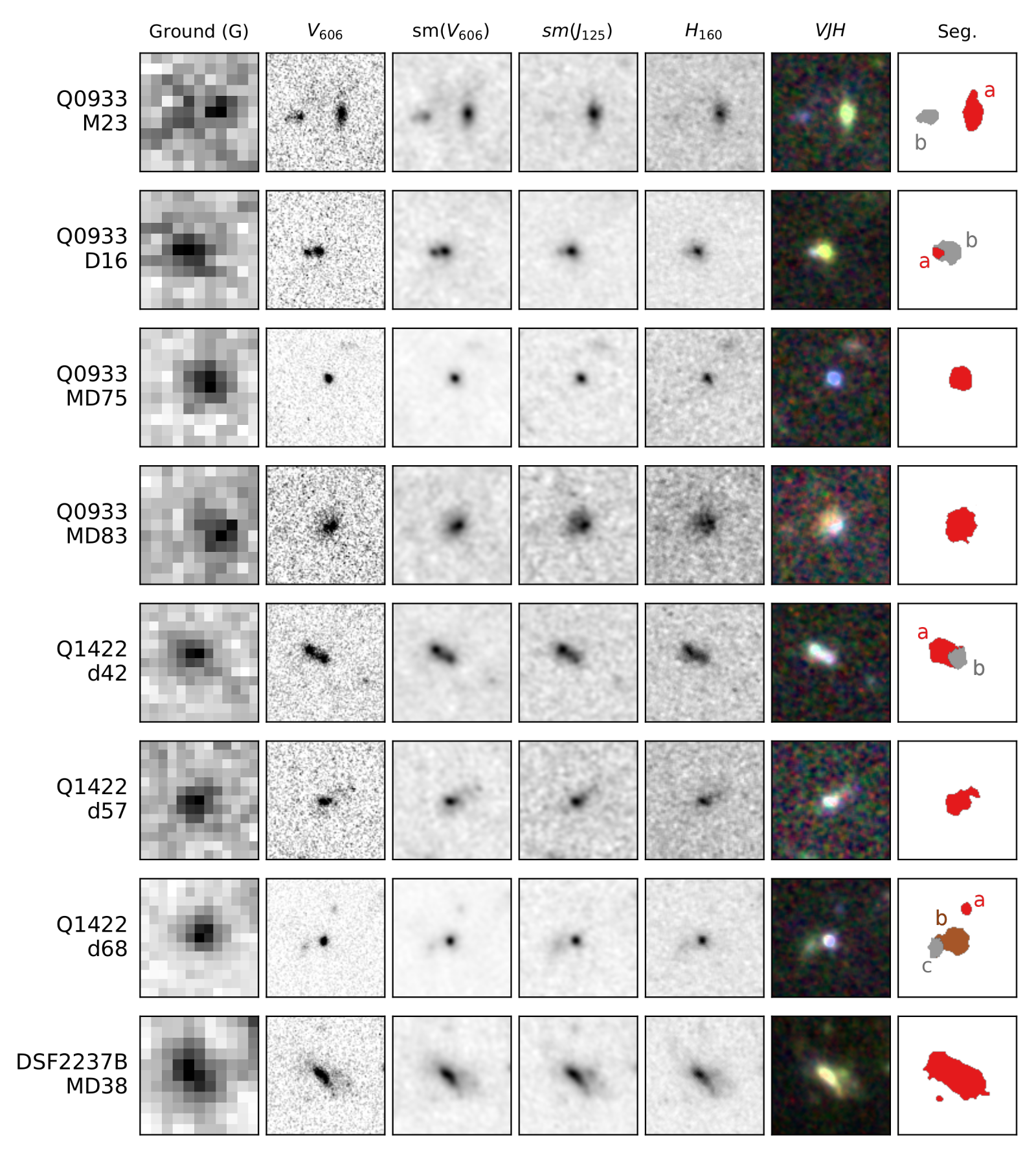}
	\caption{$3\arcsec\times3\arcsec$ postage stamps of the 15 LyC detection candidates targeted by HST. 
		\textbf{First column:} Ground-based $G$ \citep{Reddy2012}. 
		\textbf{Second column}: $V_{606}$ at the original resolution. 
		\textbf{Third and fourth columns:} $V_{125}$ and $J_{125}$ smoothed to the lower resolution of $H_{160}$. 
		\textbf{Fifth Column:} Original-resolution $H_{160}$. 
		\textbf{Sixth Column:} False-color postage stamps. The $sm(V_{606})$, $sm(J_{125})$, and $H_{160}$ images are represented by blue, green, and red, respectively. 
		\textbf{Seventh Column:} segmentation map generated by SExtractor. Separate components extracted by the program are represented by different-colored regions.
	}
	\label{fig:dets}
\end{figure*}

\begin{figure*} 
	\ContinuedFloat
	\centering
	\includegraphics[width=\textwidth]{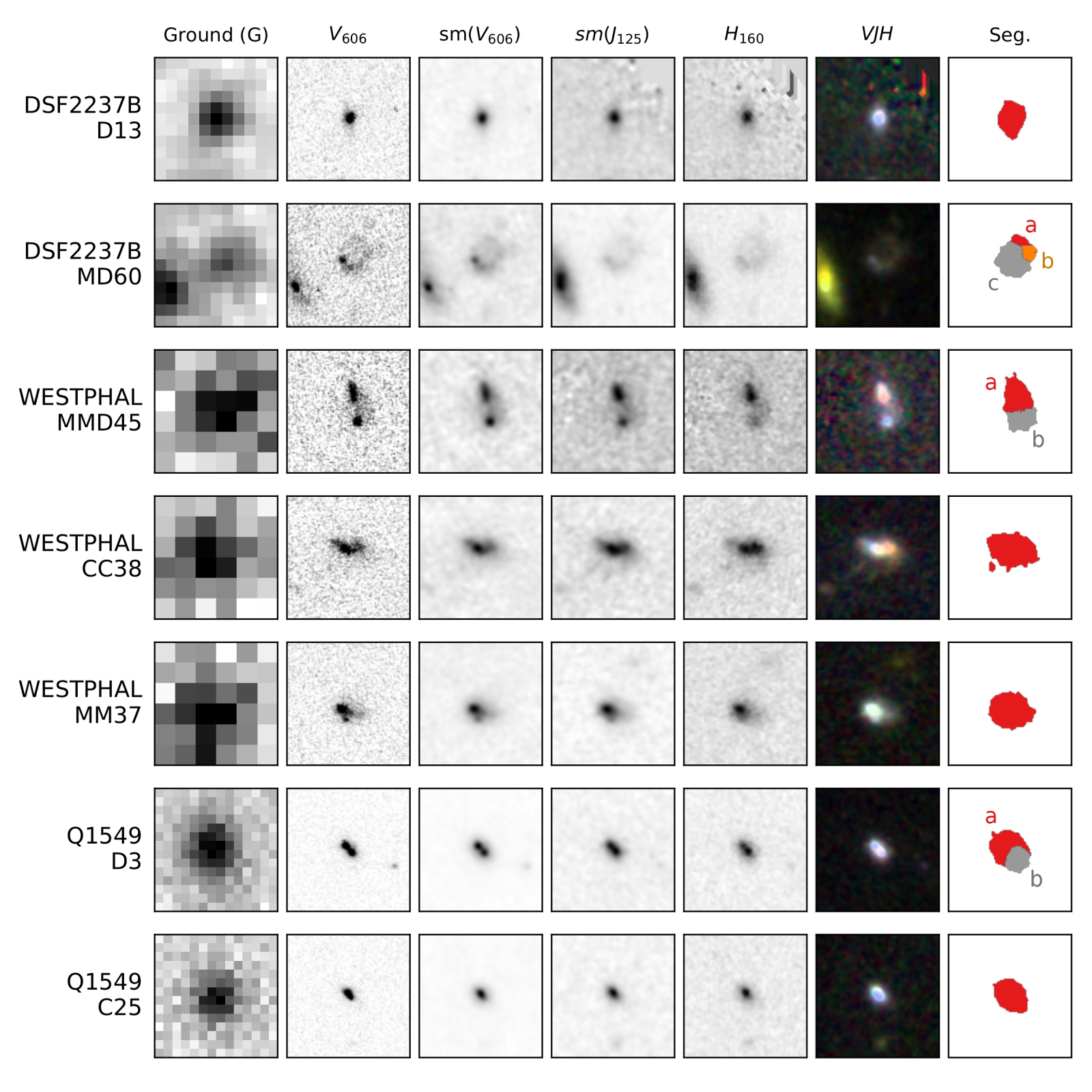}
	\caption{Continued.
	}
\end{figure*}

\subsection{Photometry} \label{sec:phot}

In order to make comparable photometric measurements of each detection in multiple filters, we first must define regions corresponding to the distinct morphological features within each object. To this end, we used $SExtractor$ on the PSF-matched $V_{606}$ image to produce segmentation maps for each object, separating the light profile into different components. We set the detection threshold at 4 times the background RMS, but reduced this parameter down to 3 times the background RMS to achieve extraction of sub-components in individual cases. For background subtraction, we used the ``local" method which uses a rectangular annulus around the source to estimate the sky level. We also set the minimum contrast parameter for deblending (\verb|DEBLEND_MINCONT|) to 0 to maximize the number of identified segments.

We placed further requirements on what could be considered a separate component for our analysis. In order to remove spurious detections, we required a minimum area of 30 pixels, corresponding to fully sampling the PSF FWHM at the resolution of $H_{160}$. We also required each component to be brighter than or equal to 28.5 mag in $V_{606}$ to eliminate all noisy and low ($<10\sigma$) S/N detections.
The segmentation maps generated for the LyC detection candidates are shown in the the last column of Figure \ref{fig:dets}. For the detection sample, eight were characterized by single component morphology, five were separated into two subcomponents, while the final two were characterized by three subcomponents. In the non-detection sample, 13 have a single component, 8 have double components, and 3 have three or more components.

For each subcomponent, we measured $V_{606}J_{125}H_{160}$ photometry from the PSF-matched mosaics. The AB photometric zeropoints of each field and filter combination were calculated using the inverse sensitivity and pivot wavelengths listed in the header of the exposures and were dust corrected using the Galactic reddening maps from \citet{Schlegel1998}. The $V_{606}$ photometry was measured isophotally by summing the $V_{606}$ flux of the smoothed image within each region in the $sm(V_{606})$ segmentation map. We used the $V_{606}$ segmentation map to measure the $J_{125}H_{160}$ photometry with \textit{SExtractor} in dual-image mode. 

We determined robust estimates of photometric error empirically. Analytical estimates of error, such as summing the error image calculated from an \textit{AstroDrizzle} exposure-time weight map, do not take into account the covariance between adjacent pixels in averaged mosaics. To reflect this covariance, \citetalias{Mostardi2015a} adopted the empirical approach of \citet{ForsterSchreiber2006}. The authors created blank apertures across the mosaics and measured the average signal RMS as a function of radius as a direct estimate of photometric errors.
This method accurately captures correlated noise in our mosaics, but doesn't take into account the spatial deviations in the error image due to differences in exposure times and sky levels.
We build upon the \citetalias{Mostardi2015a} method using the covariance correction of \citet{Law2016}.
We generated 1000 blank circular apertures in each mosaic in the range of sizes of our components with radii ranging from $0.2\arcsec$ up to $1.0\arcsec$. Within each aperture, we measured the ratio between the standard deviation of the signal within the blank aperture and the sum of the smoothed error image. This ratio is calculated as a function of angular size and indicates the correction of the theoretical error value from \textit{AstroDrizzle} to an empirical one, taking into account correlated noise in the over-sampled and smoothed mosaics.
For a given subcomponent, we take the ratio computed at the subcomponent's angular size and multiply it by the summed-in-quadrature error image within the $sm(V_{606})$ segmentation map as the 1$\sigma$ error on the photometric measurement. 
These photometric measurements for each component in the detection sample can be found in Table \ref{table:det_phot} as well as the systemic redshift ({\zspec}) from \citetalias{Steidel2018}. Descriptions of the individually-undetected sample can be found in the Appendix. The average $3\sigma$ limiting magnitudes across our pointings for the $V_{606}J_{125}H_{160}$ mosaics calculated at 1.5x the average GFWHM of the PSF were 30.18, 29.12, and 28.72 mag, respectively.

\begin{table*}
	\centering
	\caption{Photometric measurements of the LyC detection subcomponents.}
	\begin{tabular}{lcccccc}
		\toprule
		ID$^a$ &         R.A. &          Dec. & {\zspec}$^{b,c}$ &            $V_{\rm 606}$ &            $J_{\rm 125}$ &            $H_{\rm 160}$ \\
		\midrule
		DSF2237B-D13 &  22:39:27.42 &  +11:49:47.23 &         2.922 &  $24.52^{+0.03}_{-0.03}$ &  $24.95^{+0.04}_{-0.04}$ &  $24.72^{+0.04}_{-0.04}$ \\[0.10cm]
		DSF2237B-MD38 &  22:39:35.49 &  +11:50:27.50 &         3.328 &  $24.00^{+0.03}_{-0.03}$ &  $23.78^{+0.02}_{-0.02}$ &  $23.57^{+0.02}_{-0.02}$ \\[0.10cm]
		DSF2237B-MD60a &  22:39:27.91 &  +11:53:00.23 &         3.141 &  $27.25^{+0.12}_{-0.11}$ &  $27.04^{+0.08}_{-0.07}$ &  $26.68^{+0.07}_{-0.07}$ \\[0.10cm]
		DSF2237B-MD60b &  22:39:27.90 &  +11:52:59.86 &         3.141 &  $26.96^{+0.11}_{-0.10}$ &  $26.78^{+0.07}_{-0.07}$ &  $26.46^{+0.07}_{-0.06}$ \\[0.10cm]
		DSF2237B-MD60c &  22:39:27.93 &  +11:52:59.65 &         3.141 &  $24.97^{+0.05}_{-0.05}$ &  $25.00^{+0.04}_{-0.04}$ &  $24.73^{+0.04}_{-0.03}$ \\[0.10cm]
		Q0933-D16a &  09:33:30.59 &  +28:44:53.54 &         3.047 &  $27.17^{+0.06}_{-0.06}$ &  $27.03^{+0.06}_{-0.06}$ &  $26.83^{+0.07}_{-0.07}$ \\[0.10cm]
		Q0933-D16b &  09:33:30.58 &  +28:44:53.58 &           ... &  $25.90^{+0.05}_{-0.05}$ &  $25.04^{+0.02}_{-0.02}$ &  $24.88^{+0.03}_{-0.03}$ \\[0.10cm]
		Q0933-M23a &  09:33:20.62 &  +28:46:54.76 &           ... &  $25.66^{+0.06}_{-0.05}$ &  $24.99^{+0.03}_{-0.03}$ &  $24.93^{+0.05}_{-0.05}$ \\[0.10cm]
		Q0933-M23b &  09:33:20.71 &  +28:46:54.70 &         3.289 &  $26.77^{+0.09}_{-0.08}$ &  $27.57^{+0.21}_{-0.18}$ &  $27.11^{+0.22}_{-0.18}$ \\[0.10cm]
		Q0933-MD75 &  09:33:24.41 &  +28:47:46.24 &         2.913 &  $25.21^{+0.03}_{-0.03}$ &  $25.76^{+0.05}_{-0.05}$ &  $25.70^{+0.08}_{-0.08}$ \\[0.10cm]
		Q0933-MD83 &  09:33:21.51 &  +28:48:46.70 &         2.880 &  $25.42^{+0.06}_{-0.05}$ &  $25.20^{+0.05}_{-0.05}$ &  $24.88^{+0.06}_{-0.06}$ \\[0.10cm]
		Q1422-d42a &  14:24:27.75 &  +22:53:51.01 &         3.137 &  $25.47^{+0.05}_{-0.04}$ &  $25.56^{+0.05}_{-0.05}$ &  $25.45^{+0.06}_{-0.05}$ \\[0.10cm]
		Q1422-d42b &  14:24:27.73 &  +22:53:50.84 &         3.137 &  $26.04^{+0.05}_{-0.05}$ &  $26.29^{+0.07}_{-0.06}$ &  $26.07^{+0.07}_{-0.06}$ \\[0.10cm]
		Q1422-d57 &  14:24:43.25 &  +22:56:06.67 &         2.946 &  $25.68^{+0.06}_{-0.06}$ &  $25.82^{+0.08}_{-0.07}$ &  $25.65^{+0.08}_{-0.07}$ \\[0.10cm]
		Q1422-d68a &  14:24:32.92 &  +22:58:29.84 &         3.287 &  $27.76^{+0.12}_{-0.11}$ &  $28.52^{+0.30}_{-0.23}$ &  $28.50^{+0.39}_{-0.29}$ \\[0.10cm]
		Q1422-d68b &  14:24:32.94 &  +22:58:29.03 &         3.287 &  $24.86^{+0.03}_{-0.03}$ &  $25.15^{+0.04}_{-0.04}$ &  $24.91^{+0.04}_{-0.04}$ \\[0.10cm]
		Q1422-d68c &  14:24:32.98 &  +22:58:28.89 &         3.287 &  $26.81^{+0.09}_{-0.08}$ &  $26.49^{+0.07}_{-0.07}$ &  $26.65^{+0.10}_{-0.09}$ \\[0.10cm]
		Q1549-C25 &  15:52:06.07 &  +19:11:28.44 &         3.153 &  $24.55^{+0.02}_{-0.02}$ &  $24.86^{+0.03}_{-0.03}$ &  $24.68^{+0.04}_{-0.03}$ \\[0.10cm]
		Q1549-D3a &  15:51:43.72 &  +19:09:12.50 &         2.937 &  $24.65^{+0.02}_{-0.02}$ &  $24.91^{+0.03}_{-0.03}$ &  $24.67^{+0.04}_{-0.03}$ \\[0.10cm]
		Q1549-D3b &  15:51:43.71 &  +19:09:12.36 &         2.937 &  $24.96^{+0.02}_{-0.02}$ &  $25.24^{+0.03}_{-0.03}$ &  $24.87^{+0.03}_{-0.03}$ \\[0.10cm]
		Westphal-CC38 &  14:18:03.81 &  +52:29:07.23 &         3.073 &  $24.82^{+0.03}_{-0.03}$ &  $24.43^{+0.04}_{-0.04}$ &  $24.04^{+0.04}_{-0.04}$ \\[0.10cm]
		Westphal-MM37 &  14:18:26.24 &  +52:29:45.03 &         3.421 &  $24.60^{+0.02}_{-0.02}$ &  $24.31^{+0.02}_{-0.02}$ &  $24.17^{+0.02}_{-0.02}$ \\[0.10cm]
		Westphal-MMD45a &  14:18:19.81 &  +52:27:09.45 &         2.936 &  $25.64^{+0.04}_{-0.04}$ &  $25.54^{+0.05}_{-0.05}$ &  $25.04^{+0.04}_{-0.04}$ \\[0.10cm]
		Westphal-MMD45b &  14:18:19.80 &  +52:27:08.77 &         2.936 &  $26.16^{+0.05}_{-0.05}$ &  $26.18^{+0.07}_{-0.06}$ &  $25.96^{+0.08}_{-0.07}$ \\
		\bottomrule
	\end{tabular}

	\begin{flushleft}
		$^a$ {The field the object is located in, the object name, and a letter corresponding to the subcomponent in Figure \ref{fig:dets}. A subcomponent label is omitted in the case of single-component morphology.}
		$^b$ {Systemic redshift from \citetalias{Steidel2018}.}
		$^c$ {Systemic redshifts are omitted for components determined to be at low redshift based on their $V_{606}J_{125}H_{160}$ colors.}
	\end{flushleft}
\label{table:det_phot}
\end{table*}

\section{Contamination Rejection} \label{sec:contam}

As demonstrated by \citetalias{Mostardi2015a}, analysis of LyC detection candidates based on high-resolution multi-band $HST$ imaging is an effective method for identifying low-redshift interlopers. 
In \citetalias{Mostardi2015a}, photometric redshifts were estimated form the $U_{336}V_{606}J_{125}H_{160}$ magnitudes of each spatially-resolved component in the vicinity of a {\ziii} LBG.
The identification of foreground interlopers based on these component photometric redshifts led to the removal of 15 apparent LyC leakers at {\ziii} from a sample of 16. These LBGs were all photometric LyC detections that had not yet been cleaned using deep spectroscopy, as described in Section \ref{sec:sample_method} for the KLCS sample.
For the current analysis, our photometric measurements are limited to $V_{606}J_{125}H_{160}$, thus preventing us from making well-constrained photometric-redshift fits for each component. However, we are motivated to use $V_{606}J_{125}H_{160}$ photometry to predict contamination by the clear separation of low-redshift contaminants and confirmed high-redshift components of the \citetalias{Mostardi2015a} sample in $V_{606}J_{125}H_{160}$ (i.e., $V_{606} - J_{125}$ vs. $J_{125} - H_{160}$) color-color space (Figure \ref{fig:VJH_3DHST}, upper left), and the positioning of the $V_{606}J_{125}H_{160}$ filters relative to the Balmer break at {\ziii} (Section \ref{sec:sample_method}).

\subsection{$V_{606}J_{125}H_{160}$ color-color diagram} \label{sec:reject}

\begin{figure*} 
	\centering
	\includegraphics[width=\textwidth]{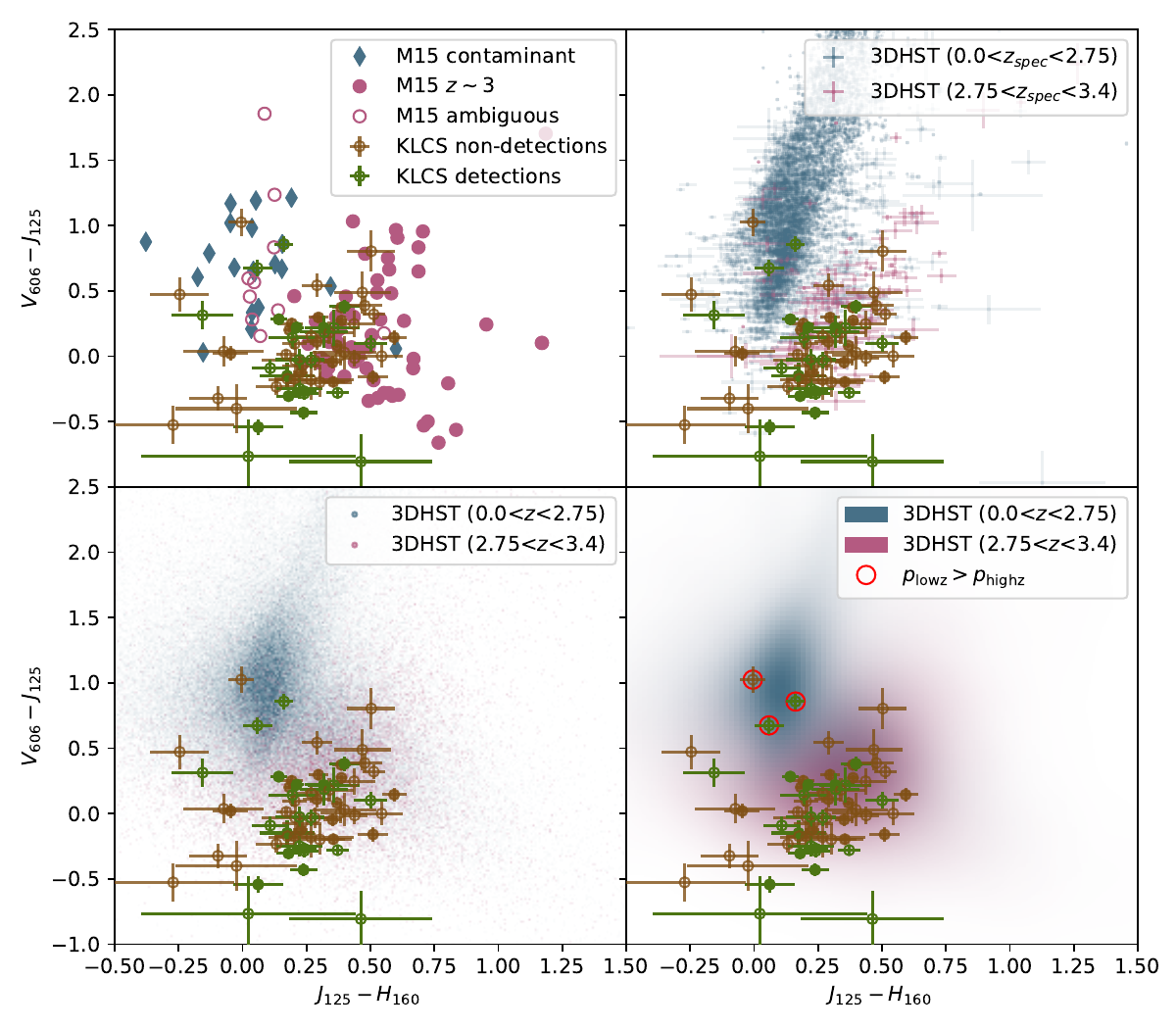}
	
	\caption{
		\textbf{Upper Left:}  $V_{606}J_{125}H_{160}$ color-color diagram of LyC detection candidates from \citetalias{Mostardi2015a}. The contaminants, selected by photometric-redshift fitting, are displayed as blue diamonds. Subcomponents predicted at {\ziii} are displayed as solid pink circles, while components with ambiguous SEDs are displayed as unfilled pink circles. Also displayed are all subcomponents extracted from the photometry of the KLCS $HST$ sample, separated into LyC detections in green and non-detections in brown. Open circles represent subcomponents from multi-component objects, while filled circles represent single-component objects.
		\textbf{Upper Right:} The subcomponents of the KLCS $HST$ sample overlaid with spectroscopic redshift samples of 3D-HST. The contaminant sample ($z_{spec}<2.75$) is displayed in blue while the {\ziii} ($2.75<z_{spec}<3.4$) sample is displayed in pink.
		\textbf{Lower Left:} The subcomponents of the KLCS $HST$ sample overlaid with photometric and spectroscopic redshift samples of 3D-HST. The redshift ranges of the two samples are identical to the spectroscopic redshift samples. Error bars of the 3D-HST samples are removed for clarity.
		\textbf{Lower Right:} The subcomponents of the KLCS $HST$ sample overlaid with the PDFs estimated from the 3D-HST samples of the lower-left panel. Each PDF was constructed using the Gaussian KDE method and were normalized to one. Components of the LyC detections Q0933-M23 and Q1422-D16 and a component of the LyC non-detection Q1422-md145 can be seen in the peak region  of the contaminant (blue) PDF and are highlighted in red.
	}
	\label{fig:VJH_3DHST}
\end{figure*}

In order to explore the validity of using the $V_{606}J_{125}H_{160}$ color-color diagram to predict foreground contamination, we examined the distribution of low- and high-redshift galaxies in $V_{606}J_{125}H_{160}$ color-color space using the photometric catalogs of the 3D-HST Survey \citep{Brammer2012,Skelton2014}. We first analysed galaxies with spectroscopically-confirmed redshifts in the redshift range of our KLCS \textit{HST} sample, $2.75<z_{spec}<3.4$, as a proxy for uncontaminated {\ziii} objects. We considered all galaxies with spectroscopic redshifts lower than our KLCS $HST$ sample, $z_{spec}<2.75$, as a representation of potential foreground contaminants. We display the low-redshift and {\ziii} populations alongside our detection and non-detection subcomponents in $V_{606}J_{125}H_{160}$ color-color space in the upper-right panel of Figure \ref{fig:VJH_3DHST}. Similar to the contaminant and {\ziii} populations of \citetalias{Mostardi2015a}, the low-redshift and {\ziii} samples of 3D-HST fall in distinct regions of $V_{606}J_{125}H_{160}$ color-color space. By using the 3D-HST sample to examine where different redshift populations of galaxies lie, we can predict whether an individual component in our analysis is a low-redshift interloper or the {\ziii} source measured by KLCS spectroscopy.

We adapted an empirical approach to rejecting low-redshift contamination in our sample of LyC detections using different 3D-HST redshift samples.
Here, we used a sample of galaxies with either photometric or spectroscopic redshifts in 3D-HST, prioritizing spectroscopic redshifts if a galaxy had both. Including galaxies with photometric redshift measurements better samples the $V_{606}J_{125}H_{160}$ dynamic range of our targets and is more complete than a spectroscopic sample weighted towards either UV selected galaxies \citep{Steidel2003,Reddy2012} or those with strong rest-optical emission lines measured through the WFC3 grism \citep{Momcheva2016,Bezanson2016}.
We again built ``contaminant" and ``{\ziii}" samples with the redshift ranges of $2.75<z<3.4$ and $z<2.75$, respectively. These 3D-HST samples are shown with the KLCS $HST$ subcomponents overlaid in the lower-left panel of Figure \ref{fig:VJH_3DHST}. 
We translated these sample distributions into a quantitative, predictive model using the Gaussian kernel-density estimation (KDE) method \citep{Parzen1962, Rosenblatt1956}. This analysis assigns each $V_{606}J_{125}H_{160}$ datapoint a 2D Gaussian profile in color-color space. These Gaussians are summed together to estimate the underlying 2D probability density function (PDF) of the distribution of data points.
This PDF can be used to estimate the likelihood that a random galaxy drawn from the sample would have a specific $V_{606}J_{125}H_{160}$.
We constructed PDFs using the KDE method for our contaminant and {\ziii} samples using the $SciPy$ routine $gaussian\_kde$ with bandwidth determined by Scott's Rule \citep{Virtanen2020, Scott2015}.
These PDFs are displayed in the lower-right panel of Figure \ref{fig:VJH_3DHST}, each normalized to one. 

We additionally examined the effects of the photometric errors of our measurements and the 3D-HST samples in the constructed PDFs.
We generated 100 realizations of the 3D-HST contaminant and {\ziii} samples randomly perturbed by their $V_{606}J_{125}H_{160}$ errors. We evaluated the resulting PDFs at the $V_{606}J_{125}H_{160}$ colors of our subcomponents, resulting in 100 probabilities from the contaminant and {\ziii} PDFs for each component. We then took the median of these probabilities and compared them on an component-to-component basis. If the median likelihood drawn from the contaminant PDF was higher than that from the {\ziii} PDF, the component was assigned as a contaminant. If the {\ziii} median likelihood was higher, the component was confirmed to be at {\ziii}.

We found that two apparent LyC detections (Q0933-M23 and Q0933-D16) contained components that had a higher likelihood of being drawn from the contaminant sample than the {\ziii} sample. We also found one LyC non-detection (Q1422-md145) that was similarly consistent with being contaminated. As a sanity check, we examined the predictions of our targets that were least likely to be contaminated, the LyC non-detections with single-component morphology. All of these objects had components consistent with being at {\ziii}. We also examined the predictions from estimating PDFs only from 3D-HST galaxies with spectroscopically-confirmed redshifts and found them to be consistent with predictions from the larger photometric-redshift 3D-HST samples.

Using the $V_{606}J_{125}H_{160}$ PDF estimated from Gaussian KDE to predict sample membership is built on a number of assumptions. By normalizing each PDF to one and comparing the probability values at a given $V_{606}J_{125}H_{160}$ directly, we assumed that each component has equal likelihood of being drawn from either distribution. We consider this assumption to be conservative, considering at least one component must be contributing spectroscopic signal consistent with {\ziii}, and the rigorous spectroscopic-blending rejection performed by \citetalias{Steidel2018}.
	
\begin{figure*} 
	\centering
	\includegraphics[width=\textwidth]{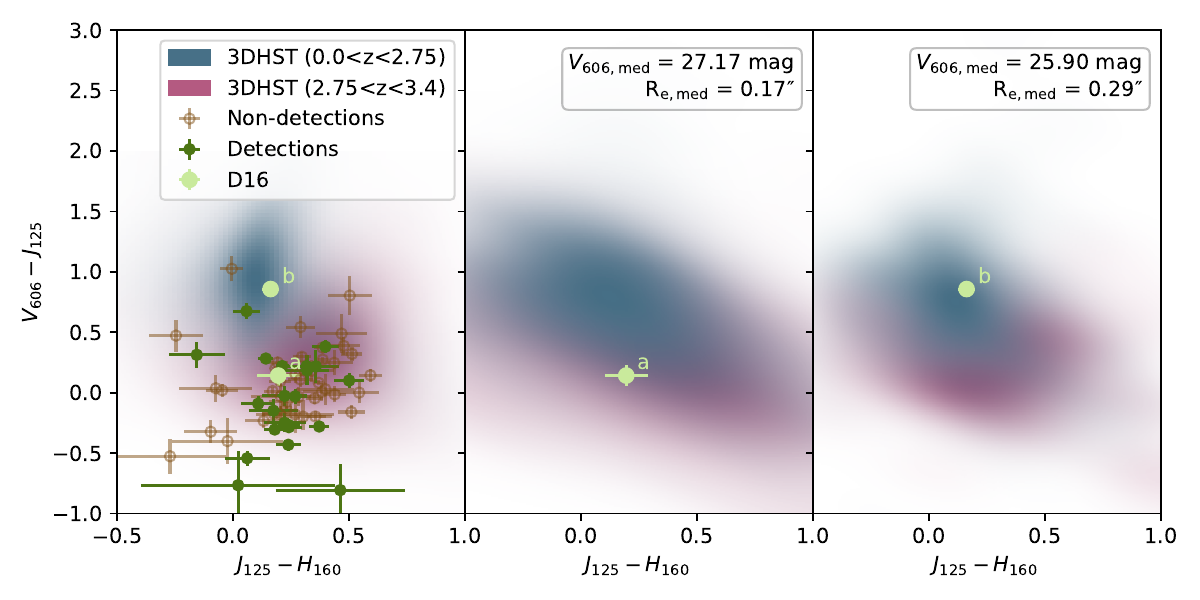}
	\caption{PDFs generated by the Gaussian KDE method on samples of 3D-HST galaxies at similar UV magnitudes and effective radii to Q0933-D16a and Q0933-D16b. 
		\textbf{Left:} PDFs generated from the full 3D-HST redshift samples in $V_{606}J_{125}H_{160}$ color-color space. The Q0933-D16a and Q0933-D16b subcomponents are highlighted. 
		\textbf{Middle and Right:} PDFs generated from 3D-HST redshift samples with similar $V_{606}$ apparent magnitude and effective radii as the subcomponents of Q0933-D16. The median $V_{606}$ apparent magnitude and $\rm R_{\rm e}$ are listed in the legend and are comparable to those of the components (see Table \ref{table:det_phot}). The same redshift ranges as the left panel were used for the low-redshift (blue) and {\ziii} (pink) distributions.
	}
	\label{fig:VJH_D16}
\end{figure*}

\begin{figure*} 
	\centering
	\includegraphics[width=\textwidth]{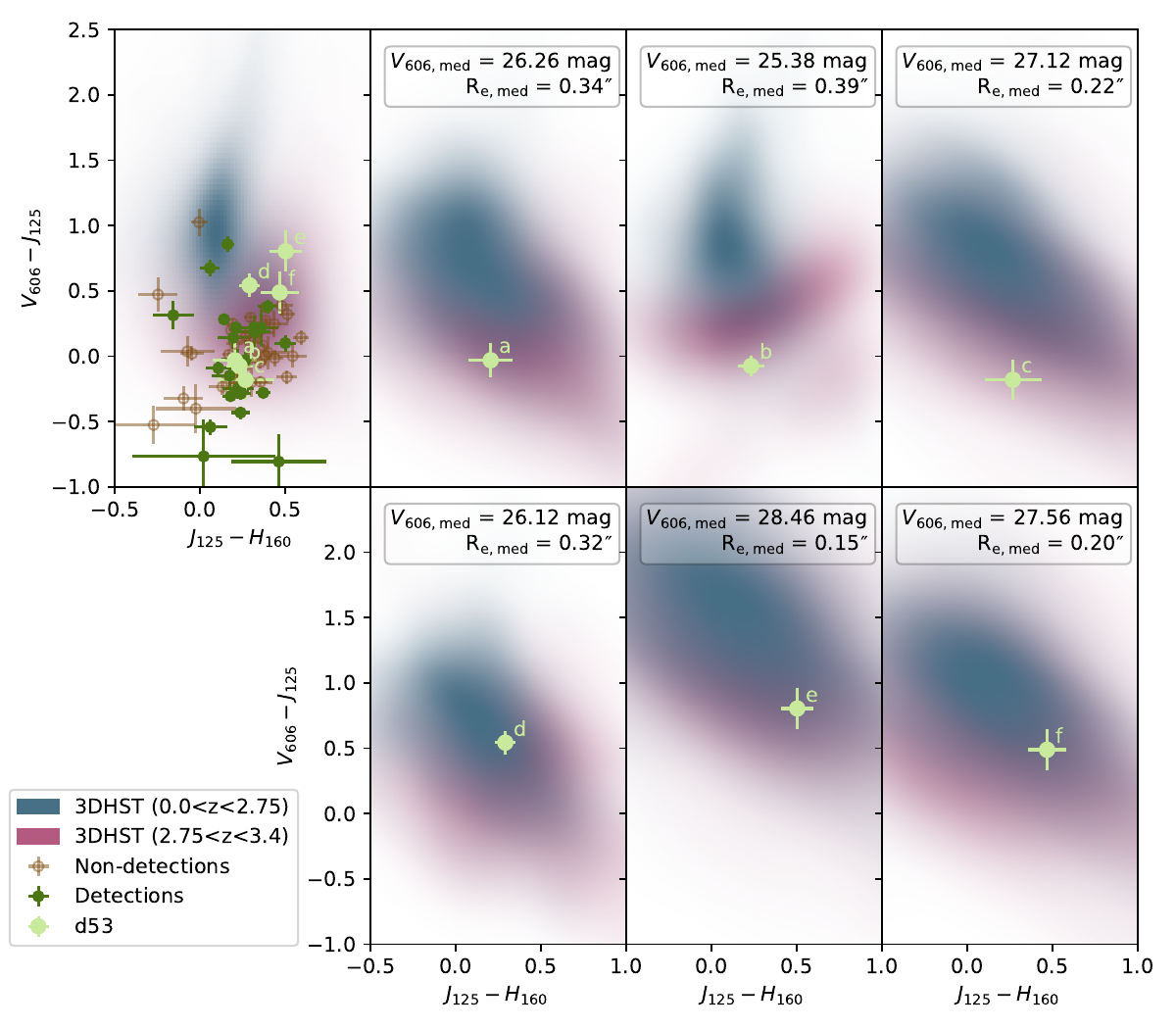}
	\caption{PDFs generated by the Gaussian KDE method on samples of 3D-HST galaxies at similar UV magnitudes and effective radii to the components of Q1422-d53. 
		\textbf{Left:} PDFs generated from the full 3D-HST redshift samples in $V_{606}J_{125}H_{160}$ color-color space. The Q1422-d53 subcomponents are highlighted. Using these PDFs, all Q1422-d53 subcomponents are predicted at {\ziii}.
		\textbf{Right:} PDFs generated from 3D-HST redshift samples with similar $V_{606}$ apparent magnitude and effective radii as the subcomponents of Q1422-d53. The median $V_{606}$ apparent magnitude and $\rm R_{\rm e}$ are listed in the legend and are comparable to those of the components (see Table \ref{table:nondet_phot}). Components \textit{d}, \textit{e}, and \textit{f} are predicted to be low redshift by these modified PDFs.
	}
	\label{fig:VJH_d53}
\end{figure*}

We also assumed that the 3D-HST contaminant and {\ziii} distributions we've constructed are representative of possible foreground contaminants and uncontaminated {\ziii} KLCS galaxies. To examine this assumption, we constructed $V_{606}J_{125}H_{160}$ PDFs that were individually tailored to the properties of each component. 
For each subcomponent, a {\ziii} and contaminant 3D-HST sample were constructed with galaxies of similar effective radii and $V_{606}$ magnitude. 3D-HST effective radii measurements were drawn from \citet{VanDerWel2012} with areas required to be within 2$\sigma$ of the size of the \textit{SExtractor} $V_{606}$ isophotal area of the subcomponent, while $V_{606}$ magnitudes were required to be within 1$\sigma$ of the subcomponent $V_{606}$ magnitude. These cuts, defined using the \citet{VanDerWel2012} errors on effective radius and our photometric-error estimate of Section \ref{sec:phot}, were set to ensure that both the {\ziii} and contaminant PDFs were well sampled. 
We bootstrapped the sample to generate median probability predictions in the method described above. 
The effects of cutting the 3D-HST samples based on effective radius and $V_{606}$ magnitude are shown for the two-component object Q0933-D16 in Figure \ref{fig:VJH_D16}. Here, we restricted the 3D-HST samples to have a median $V_{606}$ magnitude of 27.17 mag and 25.90 mag and median effective radii of 0.17$\arcsec$ and 0.29$\arcsec$ for subcomponents \textit{a} and \textit{b}, motivated by the properties of the components. 
These cuts changed the shape of the contaminant and {\ziii} PDFs, but ultimately did not affect the classification of component \textit{a} at {\ziii} and component \textit{b} as a contaminant. 
We found that the only object to be classified differently by this method was the non-detection Q1422-d53, now predicted to contain low-redshift components.
We display the 3D-HST {\ziii} and low-redshift PDFs after Q1422-d53 brightness and size cuts in Figure \ref{fig:VJH_d53}. Using these modified PDFs, components \textit{d}, \textit{e}, and \textit{f} were predicted to be low-redshift.
We considered this method to be more physically-motivated and remove Q1422-d53 from the non-detection sample.

Finally, we consider the potential effects of contamination from $2.75<z\lesssim3.0$ interlopers of galaxies with spectroscopic redshifts at the high-redshift end of our sample ($3.0<z<3.4$). Out of the 13 individual LyC detection candidates remaining once Q0933-D16 and Q0933-M23 are removed, this type of contamination is relevant for the three multicomponent galaxies, DSF2237b-MD60, Q1422-d42, and Q1422-d68 (i.e., the other 10 galaxies are either single component or at $z< 3.0$). We rebuild the 3D-HST “contaminant” sample to include higher-redshift galaxies, changing the redshift range to $z < (z_{\rm sys}-0.1)$, where $z_{\rm sys}$ is the systemic redshift of the LBG in question. Subsequently, the 3D-HST “$z\sim3$” sample was modified to have a redshift range of $(z_{\rm sys}-0.1) < z < 3.4$. Using this method, we found no difference in the classification of subcomponents  for DSF2237b-MD60, Q1422-d42, and Q1422-d68. Specifically, all subcomponent colors are consistent with being drawn from the $(z_{\rm sys}-0.1) < z < 3.4$  population.

In summary, we removed two apparent LyC detections (Q0933-D16 and Q0933-M23) and two LyC non-detections (Q1422-md145 and Q1422-d53) from the KLCS sample based on evidence of contamination from the $V_{606}J_{125}H_{160}$ colors of their morphological subcomponents.

\subsection{2D spectra}

\begin{figure} 
	\centering
	\includegraphics[width=\columnwidth]{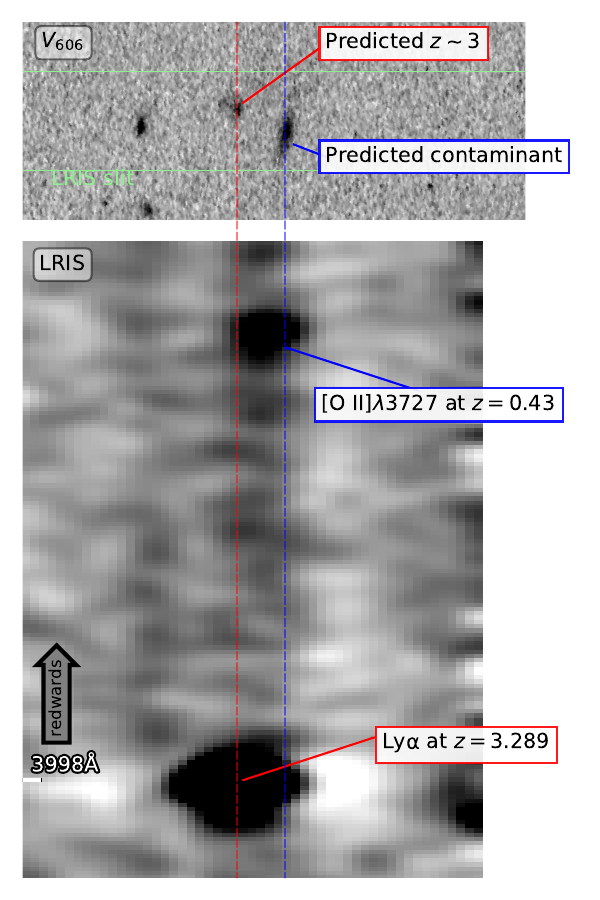}
	\caption{
		$HST\; V_{606}$ imaging and LRIS spectrum of Q0933-M23.
		\textbf{Top:} Unsmoothed $V_{606}$ postage stamp of Q0933-M23. The two components in the center of the image are associated with the extracted 1D spectrum of Q0933-M23. Color-color analysis indicates that the left component, labeled in red, is at {\ziii}, while the right component is found to be a low-redshift interloper. Displayed in light green is the location of the LRIS slit that covered this source.
		\textbf{Bottom:} 2D LRIS spectrum of Q0933-M23 oriented such that wavelength increases upwards. The horizontal position has been matched with the top panel. In the center, the blended continuum for Q0933-M23 is shown, complete with the horizontal positions of the {\ziii} and low-redshift component in dashed lines. The brighter $z=3.289$ Ly$\rm \alpha$ feature is offset towards the {\ziii} component, while a spurious feature is offset towards the low-redshift component. The continuum to the left is from an unrelated object.
	}
	\label{fig:M23_2D}
\end{figure}

In addition to examining the positions of the KLCS \textit{HST} components in $V_{606}J_{125}H_{160}$ color-color space, we analysed the high-resolution, unsmoothed $V_{606}$ images in tandem with the 2D LRIS spectra to search for new evidence of contamination. 
We examined the 2D spectra of the four objects predicted to be contaminated in Section \ref{sec:reject}. For Q0933-D16, Q1422-md145, and Q1422-d53, the subcomponents were either aligned along the slit or narrowly separated, making spectral deblending impossible. In Q0933-M23, we found evidence for foreground contamination in the LRIS spectrum shown in Figure \ref{fig:M23_2D}. Here, the Ly$\rm \alpha$ feature used for spectroscopic redshift measurement is clearly offset from the continuum towards the component predicted to lie at {\ziii} by our color-color method. There also exists an additional spectral feature offset toward the predicted low-redshift component, which was classified as Ly$\alpha$ at $z=3.380$ by \citetalias{Steidel2018}. Based on the high-resolution photometric information offered by \textit{HST} and the lack of other offset spectral features, it is more likely this emission line is [O~\textsc{ii}]$\lambda3727$ at $z=0.43$.

\section{Re-measurement of global properties} \label{sec:redo}

In this section, we revisit the results and conclusions of \citetalias{Steidel2018} with a {\ziii} rest-UV spectroscopic sample free from significant foreground contamination in the LyC spectral range. With the removal of two apparent LyC detections, Q0933-D16 and Q0933-M23, whose emission is actually contaminated by lower-redshift interlopers, global properties of the sample will be unbiased by line-of-sight contamination and will present a more accurate picture of the global ionizing properties of galaxies in the {\ziii} Universe. We also remove the LyC non-detections Q1422-md145 and Q1422-d53 due to evidence of low-redshift subcomponents within their $HST$ morphology. Our rate of significant LyC detections for individual objects ($f_{900} > 3\sigma_{900}$) has been modified from 15/124 in \citetalias{Steidel2018} to 13/124 in this work, with a foreground contamination rate of 2/15 for galaxies with individual apparent LyC detections. We detail how the removal of these contaminated objects affects sample-averaged measurements of {\fout} and {\fesca} as well as relationships between {\fout}, {\fesca} and galaxy properties. We also explore the change in ionizing emissivity in light of the cleaned sample.

Save for a small modification to the calculation of the ionizing emissivity, the methods in this section are identical to those of \citetalias{Steidel2018}.

\subsection{$\langle f_{900}/f_{1500} \rangle_{out}$} \label{sec:f900}

First, we analyse the ratio of ionizing to non-ionizing UV flux density in the uncontaminated KLCS sample. Following \citetalias{Steidel2018}, in order to average out variations of IGM transmission in different sightlines across the sky, we build a composite spectrum from the rest-frame, $f_{1500}$-normalized spectra of the sample, where $f_{1500}$ is the average flux density in the range 1475-1525~\AA{}. We combine the spectra using a  spline-interpolated common wavelength grid, computing the mean flux density at each wavelength increment after $3\sigma$ rejection of outlier flux densities. 

The ratio $\langle f_{900}/f_{1500}\rangle_{\rm obs}$ is defined as the average observed flux density of the composite spectrum in the range 880-910~\AA{}, $f_{900}$, divided by $f_{1500}$, both in units of ergs~s$^{-1}$~cm$^{-2}$~Hz$^{-1}$. 
Based on our cleaned sample of 120 galaxies, we estimate $\langle f_{900}/f_{1500}\rangle_{\rm obs}=0.015\pm0.002$, which is lower than the value of $\langle f_{900}/f_{1500}\rangle_{\rm obs}=0.019\pm0.002$ in \citetalias{Steidel2018}. However, since $f_{900}$ is heavily influenced by IGM and CGM transmission at {\ziii}, we perform transmission simulations required to estimate the ratio of ionizing to non-ionizing flux density directly outside the galaxy's ``LyC photosphere" at $r_{\rm gal}=50\:\textrm{(proper)}\:\textrm{kpc}$. This corrected ratio is defined as {\fout}. We generate 1000 estimates of $t_{\rm 900}$, the IGM+CGM transmission at 880-910~\AA{}, where each simulation comprises 120 sight lines that have an identical redshift distribution to that of our sample. 
We then divide $\langle f_{900}/f_{1500}\rangle_{\rm obs}$ by the average IGM+CGM transmission in the LyC region, $\langle t_{\rm 900}\rangle$, to obtain {\fout}, whose  uncertainty includes the uncertainty in $\langle t_{\rm 900}\rangle$.
We calculate $\langle f_{900}/f_{1500}\rangle_{\rm out}=0.040\pm0.006$ for the uncontaminated KLCS sample, compared to the ratio of $\langle f_{900}/f_{1500}\rangle_{\rm out}=0.057\pm0.006$ presented in \citetalias{Steidel2018}. This $\sim30\%$ discrepancy represents the level of contamination introduced by the strong $\sim3500$~\AA{} flux density of low-redshift interlopers in Q0933-D16 and Q0933-M23, masquerading as rest-frame LyC emission at {\ziii}.

$HST$ images are available only for the 24 LyC non-detections that lie on the pointings designed to target the apparent detection sample, thus we attempt to model the effects of potential contamination within the 85 KLCS non-detections without $HST$ imaging. We build two composite spectra. The first contains the original 24 LyC non-detections imaged by $HST$. The second removes the two galaxies with low-redshift subcomponents, Q1422-md145 and Q1422-d53, for a un-contaminated $HST$ non-detection sample of 22 galaxies. We then re-measure {\fout} from sample composite spectra in order to analyse any bias in $f_{900}$ that may arise from contamination in the full non-detection sample.  We measure {\fout} of the original non-detection $HST$ sample and the uncontaminated non-detection $HST$ sample to be $0.003\pm0.006$ and $0.002\pm0.006$, respectively. Due to the lack of significant bias in {\fout} from the inclusion of non-detections with low-redshift subcomponents, we conclude that the lack of $HST$ imaging for the larger KLCS non-detection sample will not significantly bias our global measurements of LyC flux density. Nevertheless, a larger sample of non-detections with $HST$ imaging will better quantify the level of contamination in the full sample.

We build additional composites to examine how LyC leakage correlates with a variety of galaxy properties. In addition to our full-sample composite of 120 galaxies (``All"), we build a composite with the 13 galaxies with significant $f_{900}$ (``All, detected") and a composite with the 107 galaxies with $f_{900}$ non-detections (``All, not detected"). We split the KLCS sample into four equivalent bins of increasing Ly$\rm \alpha$ equivalent width ({\wlya} (Q1-Q4)) and decreasing UV luminosity ({\luv} (Q1-Q4)). Each of these composites includes $120/4=30$ galaxies, ensuring that the uncertainty due to variation in $\langle t_{\rm 900}\rangle$ is only $\sim10\%$ \citepalias[see discussion in ][]{Steidel2018}. We also construct composites by bisecting the sample at {\wlya}$=0$ and $L_{\rm UV} = L_{\rm UV}^*$ \citep[$M_{\rm UV}=-21.0$,][]{Reddy2009} to create ``{\wlya}$>0$," ``{\wlya}$<0$," ``$L_{\rm UV} < L_{\rm UV}^*$," and ``$L_{\rm UV} > L_{\rm UV}^*$."
These composites, as well as all other composites defined by \citetalias{Steidel2018}, are reconstructed based on our cleaned sample and corrected for IGM and CGM transmission in the same manner as ``All." 
We present the {\fout} measurements for these composites in Table \ref{table:fout} along with the corresponding values from \citetalias{Steidel2018} for comparison.

\begin{table*}
	\centering
	\caption{$<f_{900}/f_{1500}>_{\rm out}$ measurements of KLCS composite spectra after contamination removal.} 
	\begin{tabular}{lcc}
		\toprule
		Sample$^a$ & $<f_{900}/f_{1500}>_{\rm out}$ (S18) & $<f_{900}/f_{1500}>_{\rm out}$ (this work) \\
		\midrule
		All &                    $0.057 \pm 0.006$ &                          $0.040 \pm 0.006$ \\
		All, detected$^b$ &                                  ... &                                        ... \\
		All, not detected$^b$ &                                  ... &                                        ... \\
		$W_{\lambda}$(Ly$\rm \alpha$) > 0 &                    $0.086 \pm 0.010$ &                          $0.063 \pm 0.009$ \\
		$W_{\lambda}$(Ly$\rm \alpha$) < 0 &                    $0.019 \pm 0.008$ &                          $0.016 \pm 0.011$ \\
		LAEs &                    $0.175 \pm 0.026$ &                          $0.107 \pm 0.023$ \\
		Non-LAEs &                    $0.032 \pm 0.008$ &                          $0.030 \pm 0.008$ \\
		$L_{UV} > L_{UV}^{*}$ &                    $0.006 \pm 0.008$ &                          $0.005 \pm 0.008$ \\
		$L_{UV} < L_{UV}^{*}$ &                    $0.113 \pm 0.014$ &                          $0.085 \pm 0.012$ \\
		$L_{UV}$ (Q1) &                    $0.005 \pm 0.008$ &                          $0.011 \pm 0.008$ \\
		$L_{UV}$ (Q2) &                    $0.000 \pm 0.011$ &                          $0.000 \pm 0.006$ \\
		$L_{UV}$ (Q3) &                    $0.114 \pm 0.018$ &                          $0.075 \pm 0.016$ \\
		$L_{UV}$ (Q4) &                    $0.138 \pm 0.024$ &                          $0.111 \pm 0.022$ \\
		z (Q1) &                    $0.053 \pm 0.018$ &                          $0.050 \pm 0.018$ \\
		z (Q4) &                    $0.056 \pm 0.011$ &                          $0.050 \pm 0.014$ \\
		$W_{\lambda}$(Ly$\rm \alpha$) (Q1) &                    $0.013 \pm 0.011$ &                          $0.005 \pm 0.010$ \\
		$W_{\lambda}$(Ly$\rm \alpha$) (Q2) &                    $0.033 \pm 0.011$ &                          $0.048 \pm 0.014$ \\
		$W_{\lambda}$(Ly$\rm \alpha$) (Q3) &                    $0.047 \pm 0.015$ &                          $0.033 \pm 0.014$ \\
		$W_{\lambda}$(Ly$\rm \alpha$) (Q4) &                    $0.166 \pm 0.025$ &                          $0.103 \pm 0.020$ \\
		$(G-R)_0$ (Q1) &                    $0.055 \pm 0.013$ &                          $0.050 \pm 0.012$ \\
		$(G-R)_0$ (Q2) &                    $0.059 \pm 0.017$ &                          $0.032 \pm 0.016$ \\
		$(G-R)_0$ (Q3) &                    $0.080 \pm 0.016$ &                          $0.089 \pm 0.017$ \\
		$(G-R)_0$ (Q4) &                    $0.029 \pm 0.016$ &                          $0.000 \pm 0.013$ \\
		\bottomrule
	\end{tabular}
	
\begin{flushleft}
	$^a$ {Full composite descriptions can be found in \citetalias{Steidel2018}.}
	$^b$ {Due to the uncertainty in the IGM+CGM correction, these entries are omitted.}
\end{flushleft}
	\label{table:fout}
\end{table*}
\subsection{Trends with {\wlya} and {\luv}}

A deep, uncontaminated spectroscopic sample covering the LyC at {\ziii} presents a unique opportunity to study how LyC leakage correlates with galaxy properties. This type of analysis is vital for understanding the physics of LyC escape and inferring a population of LyC leakers in the epoch of reionization. 

\begin{figure} 
	\centering
	\includegraphics[width=\columnwidth]{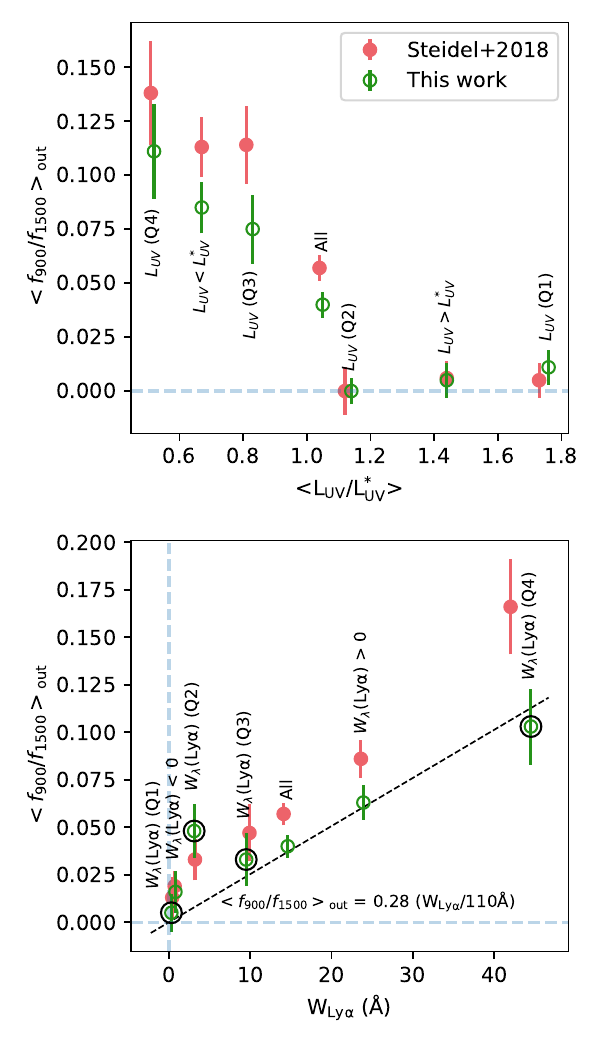}
	\caption{ Trends of {\fout} as a function of {\luv} and {\wlya} from KLCS composites. \citetalias{Steidel2018} composite measurements are shown with solid red circles, while updated measurements from this work are shown with open green circles.
		\textbf{Top:} Composites in bins of {\luv}, including four in independent quartiles of {\luv} and two bisecting the sample on $L_{\rm UV}^*$. Also displayed is the full-sample composite, ``All." A decreasing, almost bimodal relationship is seen preserved between {\fout} and {\luv}.
		\textbf{Bottom:} Composites in bins of {\wlya}, including four independent quartiles of {\wlya} and two bisecting the sample on {\wlya}$=0$, and the ``All" composite. Also displayed is the increasing, linear fit to the four independent quartiles, which are highlighted with black circles.
	}
	\label{fig:trends}
\end{figure}

We re-examine the galaxy characteristics most strongly associated with {\fout} from \citetalias{Steidel2018}: {\wlya} and {\luv}. In \citetalias{Steidel2018}, a strong correlation between {\fout} and {\wlya} was measured in the KLCS sample using composite spectra in bins of {\wlya}. In a similar manner, a strong anti-correlation between 
{\fout} and {\luv} was measured. We re-evaluate these trends using uncontaminated composites binned with respect to {\wlya} and {\luv} and present them in Figure \ref{fig:trends} alongside a measurement from the ``All" composite. 
We recover the trends of increasing {\fout} with lower {\luv} and larger {\wlya} of \citetalias{Steidel2018}, although the slopes are slightly shallower with the removal of contaminants with strong apparent $f_{900}$. 
We quantify this change in {\fout} vs. {\wlya} by performing a linear fit to the values of {\fout} in the four independent quartiles of {\wlya}, with the assumption that {\fout} tends to zero at {\wlya}$=0$. With the cleaned sample, we find 
\begin{equation} \label{eqn:fout_w}
	\langle f_{{900}}/f_{{1500}}\rangle_{{\rm out}} = 0.28 (W_{{\rm Ly\alpha}}/110\textrm{\AA}),
\end{equation}
which has a shallower slope than in the relationship $
\langle f_{{900}}/f_{{1500}}\rangle_{{\rm out}} \sim 0.36 (W_{{\rm Ly\alpha}}/110$~\AA{}) presented in \citetalias{Steidel2018}. 

These correlations imply that LyC escape occurs in galaxies with $\luv<\luv^{*}$ and strong observed Ly$\alpha$ emission relative to their UV continuum. While both relationships are compelling, we note that {\wlya} transmission has been found to be a strong function of the H~\textsc{i} covering fraction, both in the local Universe and at {\ziii} \citep[e.g., ][]{Gazagnes2018,Reddy2016}. 
To deepen our physical picture of a LyC leaking galaxy, we must consider descriptions of the structure of the neutral-phase ISM.

\subsection{\fesca}

While {\fout} is a particularly effective observational parameter for informing the global ionizing budget using only the observed far-UV (FUV) luminosity function (LF) at high-redshift \citep[e.g., \citetalias{Steidel2018};][]{Ouchi2009}, it is also important to understand how ionizing radiation escapes galaxies in relation to their intrinsic SFRs. To this end, we must simultaneously model the intrinsic UV stellar spectrum as well as the H~\textsc{i} and dust in the ISM that these photons encounter. All of these factors are required to describe the absolute escape fraction ({\fesca}), defined here as the ratio of escaping LyC photons to those produced by stars for the assumed stellar-population synthesis (SPS) model.

We estimate {\fesca} in the uncontaminated KLCS composites using SPS model fits to the IGM+CGM-corrected, composite spectra. We use identical SPS models to those in \citetalias{Steidel2018} \citep[also see ][]{Steidel2016}, with stellar metallicity $Z_*=0.001$, IMF slope $\alpha=-2.35$, and upper stellar mass limit of $300\:M_{\odot}$ from BPASS \citep[BPASS v2.1,][]{Eldridge2017}. We also assume a continuous star-formation history and an age of $t = 10^8$~yr. We allow the continuum reddening to vary across $0 \leq E(B-V) < 1.0$ for a variety of attenuation relations \citep{Reddy2015, Reddy2016, Gordon2003}. We simultaneously perform geometric ISM modeling for each composite, quantifying the covering fraction ($f_{\rm c}$) of the optically-thick H~\textsc{i} gas attenuating the FUV continuum. We adopt the ``holes" model of \citetalias{Steidel2018} to describe the physical arrangement of H~\textsc{i} and dust around the H~\textsc{ii} regions of our galaxies, where neutral H~\textsc{i} exists in a picket-fence configuration and dust is only located where neutral gas is \citep[also see][]{Reddy2016}. Accordingly, $f_{\rm c}$ varies between zero and one and the ``holes" are free of both H~\textsc{i} and dust. 
This model is consistent with correlations found between Ly$\alpha$ emission strength and low-ionization interstellar absorption strength in star-forming galaxies at intermediate redshifts \citep{Shapley2003,Steidel2010,Du2018,Pahl2020}.
\citetalias{Steidel2018} defines the attenuated observed spectrum as
\begin{equation}
	S_{\nu, \textrm{obs}}=S_{\nu, \text {int }}\left[\left(1-f_{\mathrm{c}}\right)+f_{\mathrm{c}}\: e^{-\tau}(\lambda) 10^{-A_{\lambda} / 2.5}\right],
\end{equation}
where $S_{\nu, \text {int }}$ is the intrinsic stellar spectrum, $A_{\lambda}=k_{\lambda}(E(B-V)_{\rm cov})$ where $E(B-V)_{\rm cov}$ is the continuum reddening in the foreground gas (and $k_{\lambda}$ is the wavelength-dependent dust attenuation law), and $e^{-\tau}(\lambda)$ is the transmission function due to line and continuum absorption (a function of H~\textsc{i} column density $N_{\rm HI}$). 

We fit for $E(B-V)_{\rm cov}$, log($N_{\rm HI}$) (cm$^{-2}$), and $f_{\rm c}$ for each composite and tabulate the results in Table \ref{table:fesc}. In the ``holes" model, the H~\textsc{i} and dust are optically thick to ionizing light save for the holes in the ISM where extreme UV light is transmitted entirely. Thus, the relation between {\fesca} and $f_{\rm c}$ in this ISM model is simply
\begin{equation} \label{eqn:fesc_fc}
	\centering
	f_{\rm esc,abs} = 1 - f_{\rm c}.
\end{equation}
We list the estimated {\fesca} for each composite in Table \ref{table:fesc} along with the values from \citetalias{Steidel2018} for comparison. 

A key result from this analysis is the sample-averaged {\fesca} of the KLCS survey. Based on the ``All" composite, we estimate $\fesca=0.06\pm0.01$, lower than the sample-averaged $\fesca=0.09\pm0.01$ of \citetalias{Steidel2018}. The value of {\fesca} is an essential input to models of the contribution of star-forming galaxies to reionization.
Due to increased attenuation from the neutral-phase IGM at $z>3$, observing this quantity at higher redshifts becomes increasingly difficult \citepalias{Vanzella2012,Steidel2018}. Well-constrained direct measurements of model-independent {\fout} and model-dependent {\fesca} at {\ziii} inform our picture of LyC escape at high redshift and have direct consequences for the contribution of star-forming galaxies to the ionizing background. Foreground contamination in only two galaxies in a sample of 124 can elevate {\fesca} by $\sim30\%$, underlining the need for high-resolution, multi-band imaging of LyC detection candidates at {\ziii} to vet sample-averaged measurements \citepalias{Vanzella2012, Mostardi2015a}.

We emphasize that {\fesca} is calculated here relative to BPASS SPS models that are characterized by a high ionizing photon production efficiency ($\xi_{\rm ion}=25.5$). If we instead, for example, assume an SPS model with $\xi_{\rm ion}=25.2$ \citep[e.g.,][]{Robertson2013,Robertson2015a} we would infer $\fesca$ a factor of $\sim$two larger for our full sample composite. In contrast, {\fout} is a (mostly) model-independent measure of escaping ionizing radiation that only depends on a mean IGM+CGM correction, and does not rely on assumptions about the underlying stellar population, including stellar age.

\begin{figure} 
	\centering
	\includegraphics[width=\columnwidth]{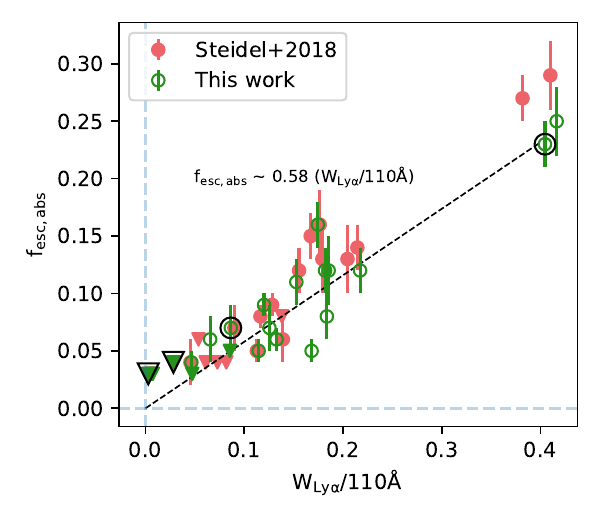}
	\caption{ Updated {\fesca} measurements for the KLCS composites as a function of {\wlya} alongside values from \citetalias{Steidel2018}. Each point represents a composite from Table \ref{table:fesc}. Triangles represent 2$\sigma$ upper limits on {\fesca}. We fit a linear trend to the four independent quartiles of {\wlya}, highlighted in black, fixing $\fesca=0$ at {\wlya}$=0$.
	}
	\label{fig:fesc_wlya}
\end{figure}

We again connect our composite measurements of LyC escape with galaxy characteristics, now in terms of the relationship between {\fesca} and {\wlya} within the uncontaminated KLCS sample. We display the {\fesca} and {\wlya} measurements for each composite in Figure \ref{fig:fesc_wlya}, including 2$\sigma$ upper limits for measurements with less than 2$\sigma$ significance. We highlight the four independent quartiles of {\wlya} in black and fit a linear trend to these points, fixing $\fesca=0$ at {\wlya}$=0$, and treating the upper limits as $\fesca=0$ with the appropriate Gaussian error. We recover the relationship
\begin{equation} \label{eqn:fesc_w}
	\centering
	f_{\rm esc,abs} = 0.58 (W_{{\rm Ly\alpha}}/110\textrm{\AA}),
\end{equation}
slightly shallower than $\fesca = 0.75 (W_{{\rm Ly\alpha}}/110$~\AA) as presented in \citetalias{Steidel2018}. This relationship between {\fesca} and {\wlya} supports the assertion that the escape of LyC photons is directly connected with the spatially-resolved distribution of H~\textsc{i} in the ISM that governs {\wlya} measurements at {\ziii} \citep[e.g., ][]{Scarlata2009,Rivera-Thorsen2015}. 
Similar relationships between {\fesca} and {\wlya} have been found in other large-scale LyC surveys at $z\sim3-4$ \citep{Marchi2017, Marchi2018, Fletcher2019}.
$f_{\rm c}$ and Ly$\alpha$ escape fraction have also been determined to predict LyC escape fractions at low redshift \citep{Chisholm2018a,Gazagnes2018}.
Additionally, $HST$ analysis of a spatially-resolved far-UV color map has shown that bluer colors are coincident with LyC escape, demonstrating that favorable H~\textsc{i} and dust configurations lead to LyC escape \citep{Ji2019}. Finally, the Ly$\alpha$ kinematics of a strongly-leaking {\ziii} galaxy have demonstrated the consistent geometry of LyC and Ly$\alpha$ escape \citep{Vanzella2020}.

\begin{table*}
	\centering
	\caption{Spectral-fitting parameters from ISM modeling using the ``holes" configuration of \citetalias{Steidel2018}.}
	\begin{tabular}{lcccccc}
		\toprule
		Sample$^a$ &  Att & $E(B-V)_{\rm cov}$ & log($N_{\rm HI
		}$) (cm$^{-2}$) & $f_c$ & {\fesca} (S18) & {\fesca} (this work) \\
		\midrule
		All &  SMC &              0.161 &               
		20.56 &  0.94 &     $0.09 \pm 0.01$ &           $0.06 \pm 0.01$ \\
		All, detected$^{b,c}$ &  SMC &              0.080 &               
		18.59 &  0.85 &     $0.31 \pm 0.03$ &           $0.21 \pm 0.03$ \\
		All, not detected &  R16 &              0.160 &               
		20.62 &  0.95 &     $0.05 \pm 0.01$ &           $0.05 \pm 0.01$ \\
		{\wlya} > 0 &  SMC &              0.056 &               
		20.12 &  0.88 &     $0.14 \pm 0.02$ &           $0.12 \pm 0.02$ \\
		{\wlya} < 0 &  R16 &              0.193 &               
		20.98 &  0.97 &               <0.03 &                     <0.03 \\
		LAEs &  SMC &              0.052 &               
		19.95 &  0.76 &     $0.29 \pm 0.03$ &           $0.25 \pm 0.03$ \\
		Non-LAEs &  R16 &              0.170 &               
		20.69 &  0.96 &     $0.04 \pm 0.02$ &           $0.04 \pm 0.01$ \\
		$L_{UV} > L_{UV}^{*}$ &  R16 &              0.166 &               
		20.72 &  0.96 &               <0.04 &                     <0.04 \\
		$L_{UV} < L_{UV}^{*}$ &  SMC &              0.064 &               
		20.38 &  0.88 &     $0.13 \pm 0.03$ &           $0.12 \pm 0.03$ \\
		$L_{UV}$ (Q1) &  R16 &              0.064 &               
		20.61 &  0.93 &               <0.04 &                     <0.05 \\
		$L_{UV}$ (Q2) &  R16 &              0.082 &               
		20.79 &  0.94 &               <0.04 &           $0.06 \pm 0.02$ \\
		$L_{UV}$ (Q3) &  SMC &              0.146 &               
		20.23 &  0.92 &     $0.13 \pm 0.03$ &           $0.08 \pm 0.02$ \\
		$L_{UV}$ (Q4) &  SMC &              0.153 &               
		20.49 &  0.89 &     $0.16 \pm 0.03$ &           $0.12 \pm 0.02$ \\
		z (Q1) &  SMC &              0.076 &               
		20.59 &  0.91 &     $0.08 \pm 0.01$ &           $0.09 \pm 0.01$ \\
		z (Q4) &  R16 &              0.098 &               
		20.36 &  0.89 &     $0.12 \pm 0.02$ &           $0.11 \pm 0.02$ \\
		{\wlya} (Q1) &  R16 &              0.185 &               
		21.06 &  0.97 &               <0.03 &                     <0.03 \\
		{\wlya} (Q2) &  R16 &              0.202 &               
		20.79 &  0.96 &               <0.04 &           $0.04 \pm 0.02$ \\
		{\wlya} (Q3) &  SMC &              0.056 &               
		19.97 &  0.93 &     $0.07 \pm 0.02$ &           $0.07 \pm 0.02$ \\
		{\wlya} (Q4) &  SMC &              0.049 &               
		20.06 &  0.77 &     $0.27 \pm 0.02$ &           $0.23 \pm 0.02$ \\
		$(G-R)_0$ (Q1) &  SMC &              0.021 &               
		20.36 &  0.84 &     $0.15 \pm 0.02$ &           $0.16 \pm 0.02$ \\
		$(G-R)_0$ (Q2) &  R16 &              0.143 &               
		20.56 &  0.93 &     $0.06 \pm 0.02$ &           $0.07 \pm 0.02$ \\
		$(G-R)_0$ (Q3) &  R16 &              0.196 &               
		20.47 &  0.95 &               <0.08 &           $0.05 \pm 0.01$ \\
		$(G-R)_0$ (Q4) &  R16 &              0.274 &               
		20.85 &  0.97 &               <0.06 &                     <0.03 \\
		\bottomrule
	\end{tabular}
\begin{flushleft}
	$^a$ {Full composite descriptions can be found in \citetalias{Steidel2018}.}
	$^b$ {Due to the unique IGM+CGM corrections described in Section \ref{sec:f900}, {\fesca} is calculated with an additional term for LyC photons traveling through an ISM with order-unity LyC optical depth.}
	$^b$ {Attenuation assumed to be in the lowest 12\%{} of the expected values.}
\end{flushleft}
	\label{table:fesc}
\end{table*}

\subsection{Ionizing emissivity and implications}

The removal of apparent LyC leakers with evidence of foreground contamination has lowered our estimates of {\fout} and {\fesca} compared to previous values based on the composites of \citetalias{Steidel2018}. Additionally, the relationships between {\fout} and {\luv},{\fout} and {\wlya}, and {\fesca} and {\wlya} have been recovered, but with slightly shallower correlation slopes.

We explore the effects of these updated results on the predictions of the galactic contributions to the ionizing background at {\ziii} by re-tracing the global emissivity calculations of \citepalias{Steidel2018}. The ionizing emissivity of galaxies can be estimated by
\begin{equation} \label{eqn:orig}
	\epsilon_{\rm LyC} \simeq 
	\int_{L_{\rm UV,min}}^{L_{\rm UV,max}} \langle f_{900}/f_{1500} \rangle_{\rm out}(L_{\rm UV}) \times L_{\rm UV} \Phi(L_{\rm UV}) dL_{\rm UV},
\end{equation}
where $\Phi(L_{\rm UV})$ is the galaxy LF evaluated in the rest-frame FUV (1500-1700~\AA{}). We can perform a simple estimate by expressing {\fout} as a function of {\luv}, taking $\langle f_{900}/f_{1500}\rangle_{\rm out}=0$ for galaxies with $M_{\rm UV}\leq -21.0$ and $\langle f_{900}/f_{1500} \rangle_{\rm out} = 0.085$ for galaxies with $-21 < M_{\rm UV}\leq -19.5$ (see ``$L_{UV} < L_{UV}^{*}$" and ``$L_{UV} > L_{UV}^{*}$" composites in Table \ref{table:fout}). Using the UV LF from \citet{Reddy2009}, we calculate a corrected $\epsilon_{\rm LyC} \simeq 9.78 \times 10^{24}$~erg~s$^{-1}$~Hz$^{-1}$~Mpc$^{-3}$, compared to the \citetalias{Steidel2018} value of $\epsilon_{\rm LyC} \simeq 13.0 \times 10^{24}$~erg~s$^{-1}$~Hz$^{-1}$~Mpc$^{-3}$.

As in \citetalias{Steidel2018}, we examine an alternate estimate of the ionizing emissivity by assuming that {\fout} varies with {\wlya}. This estimation is supported by the strong relationship between the two properties in Figure \ref{fig:trends} and can be extrapolated to higher redshifts if {\wlya} is corrected for IGM opacity. 
For this calculation, we move {\fout} outside the integral in Equation \ref{eqn:orig}, and estimate an average {\fout} weighted by {\wlya}. We multiply this average {\fout}, E({\fout}), by the integral over the non-ionizing UV LF, $\epsilon_{\rm UV}$.
To find the {\wlya}-weighted average value of {\fout}, we combine our relationships between {\fout} and {\wlya} in Equation \ref{eqn:fout_w} and the distribution function $n($\wlya$)$.

While it is possible to combine Equation \ref{eqn:fout_w} and $n($\wlya$)$ directly, we introduce physical descriptions of Ly$\alpha$ escape to extrapolate the behavior of {\fout} outside of the range of {\wlya} in our sample. \citetalias{Steidel2018} connected the modulation of {\wlya} by $f_{\rm c}$ through two separate processes: the scattering of Ly$\alpha$ photons by foreground gas and the decrease in the Ly$\alpha$ source function when H~\textsc{ii} regions are no longer optically thick to ionizing radiation. These processes connect through the equation
\begin{equation} \label{eqn:}
	(1-f_{\rm c}) = 0.5 - [|0.5W' - 0.25|]^{1/2},
\end{equation}
where $W'\equiv $\wlya$/W_{\lambda}$(Ly$\rm \alpha$, Case B) and $W_{\lambda}$(Ly$\rm \alpha$, Case B)=110~\AA{} in our assumed SPS model. We can join this physically-motivated parameterization with the empirical relations in Equations \ref{eqn:fout_w} and \ref{eqn:fesc_w} to express {\fout} in terms of $W'$:  
\begin{equation} \label{eqn:fout_calc}
	\langle f_{900}/f_{1500}\rangle_{\rm out} \sim 
	0.24 - [|0.117W' - 0.058|]^{1/2}.
\end{equation}
To calculate the expectation value of {\fout} weighted by $W'$, we now include the relative incidence of {\wlya}, 
\begin{equation} \label{eqn:n}
	n(W_{\lambda}) \propto \textrm{exp}(-W_{\lambda}/23.5\textrm{\AA}),
\end{equation}
consistent with \citetalias{Steidel2018} and spectroscopic samples analysed in \citet{Shapley2003} and \citet{Kornei2010}. To find the expectation value of {\fout}, E({\fout}), we simplify the method of \citetalias{Steidel2018} by integrating Equation \ref{eqn:n} with a change of variables from Equation \ref{eqn:fout_calc}. After including an assumption that 40$\%$ of galaxies have \wlya$\leq0$ and therefore \fout$\simeq0$ for such sources, we find that the integral becomes
\begin{equation}
	\begin{aligned}
	E(\langle f_{900}/f_{1500}\rangle_{\rm out}) = \dfrac{\splitdfrac{0.6 \int_{0}^{1} [0.24 - [|0.117W' - 0.058|]^{1/2}]}
		{\times \textrm{exp}(-W'\times(110 / 23.5)) dW'}}{\int_{0}^{1} exp(-W'\times(110 / 23.5)) dW'} 
	\end{aligned}
\end{equation}
and evaluates to $E(\langle f_{900}/f_{1500}\rangle_{\rm out}) = 0.032$. Based on the equivalent values relating {\fout}, {\fesca}, and {\wlya} from \citetalias{Steidel2018}, this method estimates $E(\langle f_{900}/f_{1500}\rangle_{\rm out}) = 0.042$. We then multiply $E(\langle f_{900}/f_{1500}\rangle_{\rm out})$ by $\epsilon_{\rm UV}$ to find $\epsilon_{\rm LyC} \simeq 5.5 \times 10^{24}$~erg~s$^{-1}$~Hz$^{-1}$~Mpc$^{-3}$. For comparison, when we apply this methodology to the contaminated \citetalias{Steidel2018} results, we find a higher emissivity of $\epsilon_{\rm LyC} \simeq 7.2 \times 10^{24}$~erg~s$^{-1}$~Hz$^{-1}$~Mpc$^{-3}$. Our value remains consistent with other recent estimates, such as those in \citet{Jones2021}: $\epsilon_{\rm LyC} \simeq 10.0_{-5.0}^{+10.0} \times 10^{24}$~erg~s$^{-1}$~Hz$^{-1}$~Mpc$^{-3}$ at $z\sim2.5$ and $\epsilon_{\rm LyC} \simeq 15.8_{-10.8}^{+34.3} \times 10^{24}$~erg~s$^{-1}$~Hz$^{-1}$~Mpc$^{-3}$  at $z\sim2.9$.

The reduction of ionizing emissivity from star-forming galaxies found using the uncontaminated composites of this work is significant. However, galaxies still provide a comparable contribution to the ionizing background at {\ziii} to that of active galactic nuclei (AGNs), which are estimated to contribute between 1.6 and $10\times10^{24}$~erg~s$^{-1}$~Hz$^{-1}$~Mpc$^{-3}$ at these redshifts \citep{Hopkins2007,Cowie2009,Kulkarni2019,Shen2020}. 
Furthermore, the unique size, purity, and data products of the sample in this work can inform the details of reionization models and the physics of LyC escape at $z>3$. Authors have recently described competing models of reionization as  ``democratic," where reionization is driven by numerous faint sources with high {\fesca} \citep[e.g., ][]{Finkelstein2019}, and ``oligarchical," where $<$$5\%$ of galaxies contribute $>$$80\%$ of the reionization budget \citep{Naidu2020}. Both of these interpretations depend on base assumptions of the evolution and value of {\fesca} across cosmic time and the degree of LyC leakage associated with a range of galaxy properties. 

\citet{Finkelstein2019} used a halo-mass dependent {\fesca} in their fiducial  model. These authors also considered a fixed $\fesca=0.09\pm0.01$, as found in \citetalias{Steidel2018}, and accordingly report tension with observational constraints on the progress of reionization.
However, in order to consider the implications of different {\fesca} values on the evolution of the ionizing emissivity, careful attention must be paid to the manner in which key quantities are empirically estimated, modeled, or assumed. Given that the most robust quantity we estimate is {\fout}, the only fair way to incorporate our {\fesca} value into the \citet{Finkelstein2019} model is also to assume the same intrinsic LyC luminosity for similar galaxies.
Additionally, the fiducial model of \citet{Finkelstein2019} predicts that AGNs dominate the ionizing budget even extrapolated to {\ziii}, a determination in tension with this work and \citetalias{Steidel2018}.
Keeping in mind the same caveats about incorporating an {\fesca} determined from observations into reioniziation models, we turn to \citet{Naidu2020}, who built two empirical models, one that fits for a constant $\fesca=0.21_{-0.04}^{+0.06}$ during reionization and one that parameterizes {\fesca} as a function of SFR surface density ($\Sigma_{\rm SFR}$). 
For the second model specifically, these authors include $f_{\rm esc,abs}=0.09\pm0.01$ from \citetalias{Steidel2018} as a constraint in the fit, and predict $\sim10\%$ of sources have $\fesca>20$\% at $z\sim4$. The \citeauthor{Naidu2020} model would have to be tweaked further to incorporate a lower average {\fesca} at {\ziii}, but the rough percentage of sources with large escape fractions remains consistent with the KLCS detection fraction. This model is also built upon the assumption that the leakage of LyC is correlated with increasing $\Sigma_{\rm SFR}$, a determination that can now be tested with the KLCS sample using dust-corrected SFR values and $HST$ sizes. In future work, we will focus on the properties of the LyC leaking galaxies in the uncontaminated KLCS survey, including stellar mass, $E(B-V)$, $\Sigma_{\rm SFR}$, and age.

\section{Summary} \label{sec:summary}

Due to the opacity of the IGM to ionizing photons for $z>3.5$, direct detections of the LyC in galaxies at lower redshifts are vital for drawing inferences about the ionizing spectra of reionization-era galaxies. 
With the goal of increasing the number of confirmed individual LyC detections and removing contamination from sample-averaged measurements at {\ziii}, we examined LyC detection candidates of the KLCS survey for foreground contamination. By testing each morphological sub-component for low-redshift signatures with the $V_{606}J_{125}H_{160}$ color-color diagram, we removed contaminated objects from the KLCS sample and revisited the measurements of \citetalias{Steidel2018} after building un-contaminated composite spectra. 
The key results are as follows:
\begin{enumerate}
	\item We find evidence of foreground contamination in the subcomponents of two LyC detection candidates, Q0933-M23 and Q0933-D16, as well as two non-detections, Q1422-d53 and Q1422-md145. The fraction of contaminated sources within the apparent LyC detections was 2/15, and the subsequent removal of these sources brought the LyC detection fraction for individual objects of the KLCS survey from 15/124 to 13/124.
	\item We re-measured the sample-averaged ratio of ionizing to non-ionizing rest-UV flux density after performing corrections for IGM and CGM attenuation, finding $\fout=0.040\pm0.006$. This value has been reduced by $\sim$30\% compared to the original $\fout=0.057\pm0.006$ of \citetalias{Steidel2018}, indicating the contribution of low-redshift flux density to the KLCS composite spectra. This significant reduction underscores the necessity for high-resolution, multi-band imaging to remove bias in {\ziii} sample-averaged quantities related to the LyC spectrum.
	\item The strong dependence of {\fout} on {\wlya} and {\luv} within the KLCS sample remains in the contamination-corrected sample. {\fout} increased from $0.005\pm0.008$ to $0.085\pm0.014$ in composites built from the UV-brightest and dimmest halves of the KLCS sample. We also measured a positive, linear correlation between {\fout} and {\wlya} from composite spectra, tying together the escape of LyC and Ly$\alpha$ photons in the ISM of our galaxies.
	\item We estimated a sample-averaged absolute escape fraction of $\fesca=0.06\pm0.01$  at {\ziii} after performing stellar-population fits and ISM modeling of the rest-UV spectrum. Similar to {\fout}, this uncontaminated value is lower than the \citetalias{Steidel2018} {\fesca} of $0.09\pm0.01$ for the same assumptions concerning the SPS model. We also recovered the positive, linear relationship between {\fesca} and {\wlya}. This relationship can be interpreted as the H~\textsc{i} covering fraction modulating the strengths of both Ly$\alpha$ emission and LyC leakage, supported by relationships between {\wlya} and $f_{\rm c}$ that have been found locally \citep[e.g., ][]{Gazagnes2018}.
	\item Based on our modified results, we re-estimate the ionizing emissivity of $L_{UV} > 0.3L_{UV}^{*}$ galaxies at {\ziii}, determining $\epsilon_{\rm LyC} \simeq 5.5 \times 10^{24}$~erg~s$^{-1}$~Hz$^{-1}$~Mpc$^{-3}$. We perform this calculation using the observed functional dependence of {\fout} on {\wlya}, the {\wlya} distribution function, and the far-UV luminosity function. Based on this uncontaminated estimate, we conclude that the ionizing background at {\ziii} contains comparable contributions of both galaxies and AGNs.
\end{enumerate}
These results have important implications for models of reionization.
Mapping the evolution of the IGM neutral fraction to the change in galaxy populations with redshift requires assumptions of {\fesca} that must be consistent with direct, sample-averaged results such as those from this work. Moreover, the degree to which {\fesca} and {\fout} can be affected by just a few low-redshift interlopers demonstrates the utility of high-resolution follow-up of any individual or statistical detection of {\ziii} LyC. As the sample of confirmed LyC detections widens through deep, spectroscopic and narrow-band imaging surveys, our understanding of the ionizing characteristics of reionization-era galaxies will follow.
\linebreak

Support for program HST‐GO‐15287.001 was provided by NASA through a grant from the Space Telescope Science Institute, which is operated by the Associations of Universities for Research in Astronomy, Incorporated, under NASA contract NAS5-26555. 
CS and YC were supported in part by the Caltech/JPL President's and Director's program.
We wish to extend special thanks to those of Hawaiian ancestry on
whose sacred mountain we are privileged to be guests. Without their generous hospitality, most
of the observations presented herein would not have been possible.

\section*{Data Availability Statement}
The {\it HST} data presented in this article are publicly available
from the Mikulski Archive for Space Telescopes. The ground-based
data presented here will be shared on reasonable request to the
corresponding author.


\appendix
\section{Non-detections}

Here we describe the properties of the 24 LyC non-detections targeted by \textit{HST}. In Figure \ref{fig:nondets}, we show the $V_{606}J_{125}H_{160}$ postage stamps of each non-detection alongside ground-based images, \textit{HST} false-color images, and segmentation maps in a manner similar to Figure \ref{fig:dets}. We report the photometric measurements of each subcomponent displayed in Figure \ref{fig:nondets} in Table \ref{table:nondet_phot}.

\begin{figure*} 
	\centering
	\includegraphics[width=\textwidth]{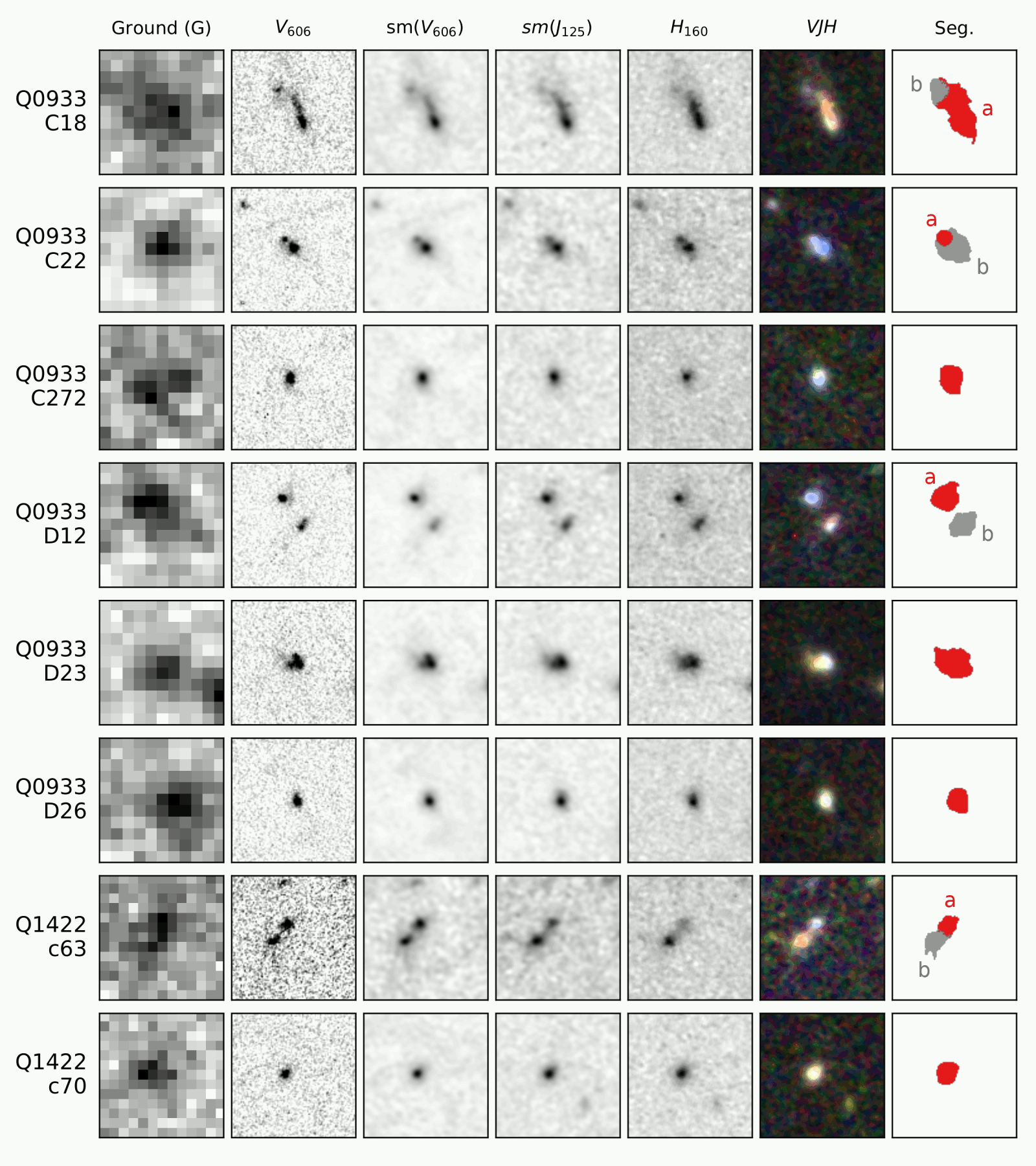}
	\caption{$3\arcsec\times3\arcsec$ postage stamps of the 24 LyC non-detections targeted by HST.
		\textbf{First column:} Ground-based $G$ \citep{Steidel2003}. 
		\textbf{Second column}: $V_{606}$ at the original resolution. 
		\textbf{Third and fourth columns:} $V_{125}$ and $J_{125}$ smoothed to the lower-resolution of $H_{160}$. 
		\textbf{Fifth Column:} Original-resolution $H_{160}$. 
		\textbf{Sixth Column:} False-color postage stamps. The $sm(V_{606})$, $sm(J_{125})$, and $H_{160}$ images are represented by blue, green, and red, respectively. 
		\textbf{Seventh Column:} segmentation map generated by SExtractor. Separate components extracted by the program are represented by different-colored regions.
	}
	\label{fig:nondets}
\end{figure*}

\begin{figure*} 
	\ContinuedFloat
	\centering
	\includegraphics[width=\textwidth]{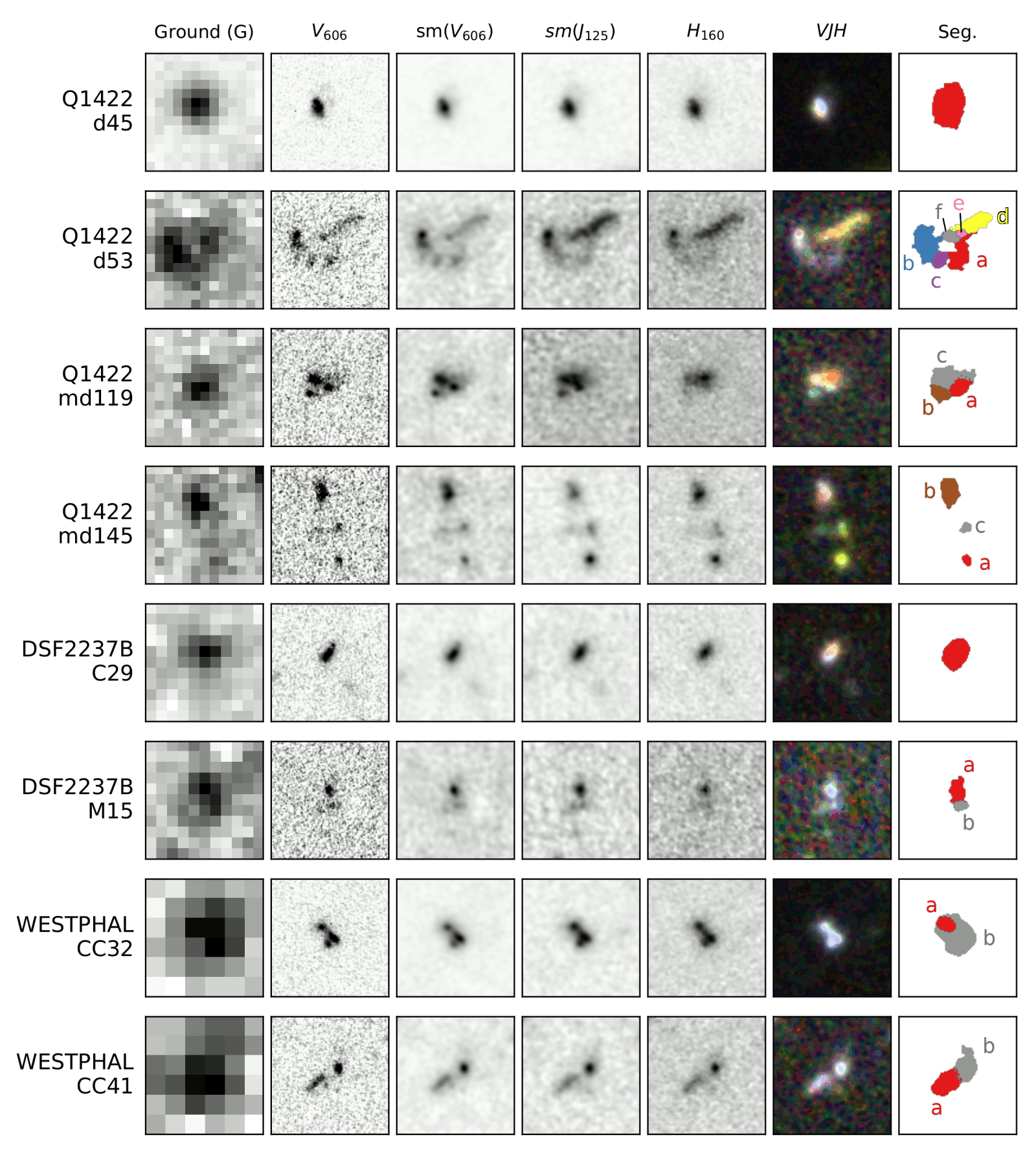}
	\caption{Continued.
	}
\end{figure*}

\begin{figure*} 
	\ContinuedFloat
	\centering
	\includegraphics[width=\textwidth]{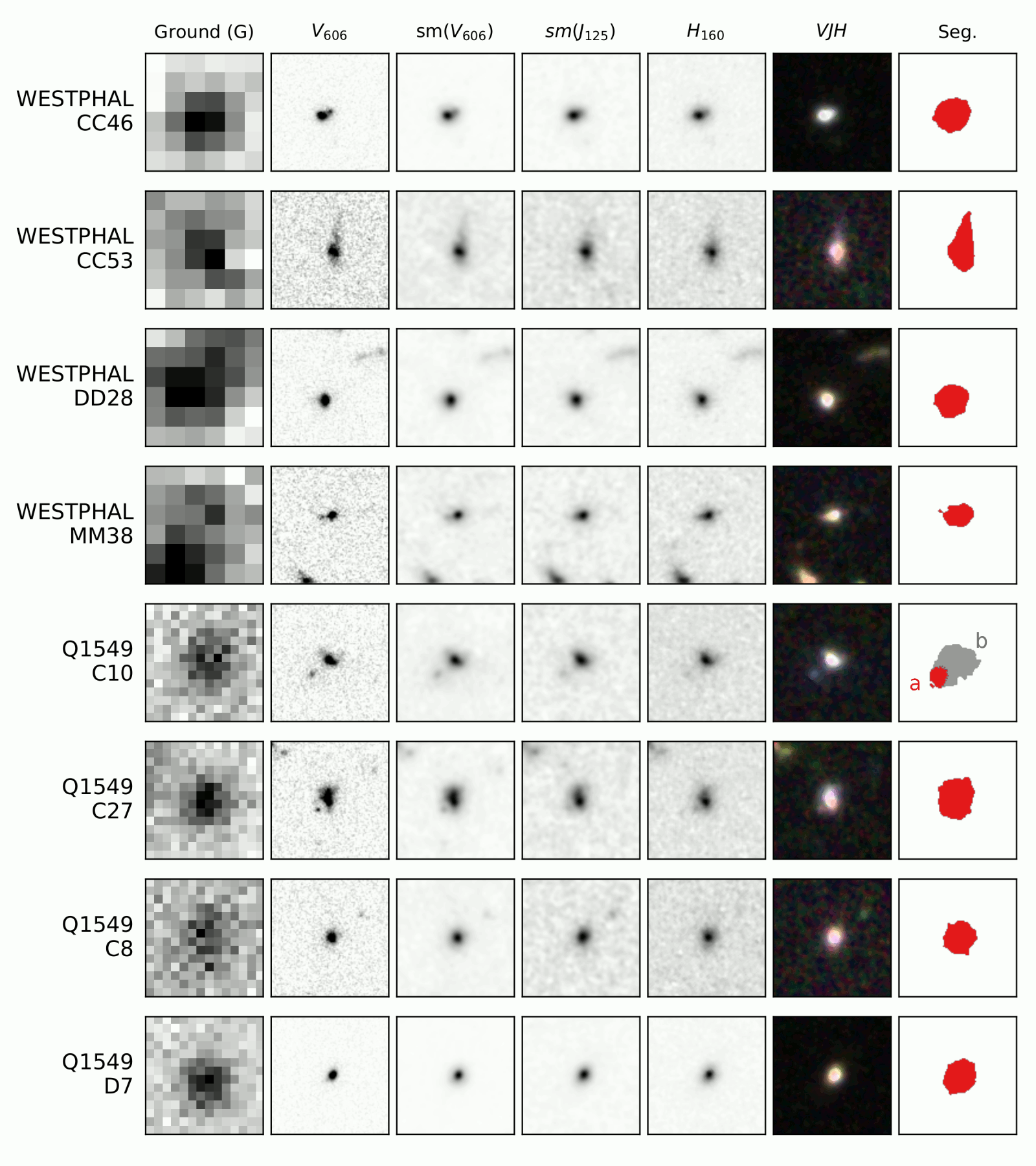}
	\caption{Continued.
	}
\end{figure*}

\begin{table*}
	\centering
	\caption{Photometric measurements of the LyC non-detection subcomponents.}
	\begin{tabular}{lcccccc}
		\toprule
		ID$^a$ &         R.A. &          Dec. & {\zspec}$^{b,c}$ &            $V_{\rm 606}$ &            $J_{\rm 125}$ &            $H_{\rm 160}$ \\
		\midrule
		DSF2237B-C29 &  22:39:34.50 &  +11:52:41.15 &         3.100 &  $24.83^{+0.04}_{-0.04}$ &  $24.88^{+0.03}_{-0.03}$ &  $24.53^{+0.03}_{-0.02}$ \\[0.10cm]
		DSF2237B-M15a &  22:39:30.71 &  +11:51:40.93 &         3.403 &  $26.02^{+0.07}_{-0.06}$ &  $26.34^{+0.07}_{-0.06}$ &  $26.44^{+0.10}_{-0.09}$ \\[0.10cm]
		DSF2237B-M15b &  22:39:30.71 &  +11:51:40.57 &         3.403 &  $27.45^{+0.12}_{-0.11}$ &  $27.85^{+0.16}_{-0.14}$ &  $27.87^{+0.20}_{-0.17}$ \\[0.10cm]
		Q0933-C18a &  09:33:34.24 &  +28:43:23.39 &         2.926 &  $24.89^{+0.05}_{-0.05}$ &  $24.57^{+0.04}_{-0.03}$ &  $24.06^{+0.03}_{-0.03}$ \\[0.10cm]
		Q0933-C18b &  09:33:34.28 &  +28:43:23.98 &         2.926 &  $26.69^{+0.09}_{-0.09}$ &  $26.66^{+0.10}_{-0.09}$ &  $26.26^{+0.09}_{-0.08}$ \\[0.10cm]
		Q0933-C22a &  09:33:32.12 &  +28:43:42.26 &         3.164 &  $26.53^{+0.34}_{-0.26}$ &  $26.56^{+0.48}_{-0.33}$ &  $26.28^{+0.51}_{-0.35}$ \\[0.10cm]
		Q0933-C22b &  09:33:32.10 &  +28:43:42.08 &         3.164 &  $25.21^{+0.03}_{-0.03}$ &  $25.44^{+0.04}_{-0.04}$ &  $25.31^{+0.05}_{-0.05}$ \\[0.10cm]
		Q0933-C272 &  09:33:27.30 &  +28:44:37.37 &         3.546 &  $25.42^{+0.04}_{-0.04}$ &  $25.40^{+0.04}_{-0.04}$ &  $25.45^{+0.06}_{-0.06}$ \\[0.10cm]
		Q0933-D12a &  09:33:33.95 &  +28:44:18.83 &         2.924 &  $25.38^{+0.04}_{-0.04}$ &  $25.56^{+0.05}_{-0.05}$ &  $25.36^{+0.06}_{-0.06}$ \\[0.10cm]
		Q0933-D12b &  09:33:33.92 &  +28:44:18.19 &         2.924 &  $25.95^{+0.06}_{-0.06}$ &  $25.95^{+0.07}_{-0.06}$ &  $25.40^{+0.06}_{-0.05}$ \\[0.10cm]
		Q0933-D23 &  09:33:23.44 &  +28:47:17.11 &         3.224 &  $24.71^{+0.03}_{-0.03}$ &  $24.41^{+0.02}_{-0.02}$ &  $24.12^{+0.03}_{-0.03}$ \\[0.10cm]
		Q0933-D26 &  09:33:25.08 &  +28:48:24.41 &         3.266 &  $25.34^{+0.03}_{-0.03}$ &  $25.09^{+0.03}_{-0.03}$ &  $24.90^{+0.04}_{-0.03}$ \\[0.10cm]
		Q1422-c63a &  14:24:30.19 &  +22:53:56.52 &         3.059 &  $26.51^{+0.07}_{-0.07}$ &  $26.70^{+0.09}_{-0.09}$ &  $26.40^{+0.09}_{-0.08}$ \\[0.10cm]
		Q1422-c63b &  14:24:30.22 &  +22:53:56.14 &         3.059 &  $26.42^{+0.08}_{-0.08}$ &  $26.18^{+0.07}_{-0.06}$ &  $25.74^{+0.05}_{-0.05}$ \\[0.10cm]
		Q1422-c70 &  14:24:33.65 &  +22:54:55.27 &         3.129 &  $25.72^{+0.05}_{-0.05}$ &  $25.61^{+0.04}_{-0.04}$ &  $25.32^{+0.04}_{-0.04}$ \\[0.10cm]
		Q1422-d45 &  14:24:32.23 &  +22:54:03.16 &         3.072 &  $23.77^{+0.02}_{-0.02}$ &  $23.90^{+0.02}_{-0.02}$ &  $23.67^{+0.02}_{-0.02}$ \\[0.10cm]
		Q1422-d53a &  14:24:25.53 &  +22:55:00.28 &         3.086 &  $26.25^{+0.10}_{-0.09}$ &  $26.28^{+0.09}_{-0.09}$ &  $26.08^{+0.10}_{-0.09}$ \\[0.10cm]
		Q1422-d53b &  14:24:25.59 &  +22:55:00.71 &         3.086 &  $25.38^{+0.06}_{-0.05}$ &  $25.45^{+0.06}_{-0.05}$ &  $25.22^{+0.06}_{-0.05}$ \\[0.10cm]
		Q1422-d53c &  14:24:25.57 &  +22:55:00.24 &         3.086 &  $27.13^{+0.11}_{-0.10}$ &  $27.31^{+0.12}_{-0.11}$ &  $27.04^{+0.13}_{-0.11}$ \\[0.10cm]
		Q1422-d53d &  14:24:25.50 &  +22:55:01.22 &           ... &  $26.12^{+0.08}_{-0.07}$ &  $25.57^{+0.05}_{-0.04}$ &  $25.28^{+0.04}_{-0.04}$ \\[0.10cm]
		Q1422-d53e &  14:24:25.53 &  +22:55:00.92 &           ... &  $28.48^{+0.15}_{-0.13}$ &  $27.67^{+0.07}_{-0.07}$ &  $27.17^{+0.06}_{-0.06}$ \\[0.10cm]
		Q1422-d53f &  14:24:25.54 &  +22:55:00.83 &           ... &  $27.56^{+0.14}_{-0.13}$ &  $27.07^{+0.09}_{-0.08}$ &  $26.61^{+0.07}_{-0.07}$ \\[0.10cm]
		Q1422-md119a &  14:24:36.19 &  +22:55:40.33 &         2.751 &  $26.44^{+0.06}_{-0.05}$ &  $26.26^{+0.07}_{-0.07}$ &  $25.92^{+0.07}_{-0.06}$ \\[0.10cm]
		Q1422-md119b &  14:24:36.22 &  +22:55:40.17 &         2.751 &  $26.74^{+0.06}_{-0.06}$ &  $26.70^{+0.10}_{-0.09}$ &  $26.78^{+0.13}_{-0.12}$ \\[0.10cm]
		Q1422-md119c &  14:24:36.21 &  +22:55:40.50 &         2.751 &  $25.73^{+0.05}_{-0.05}$ &  $25.34^{+0.05}_{-0.05}$ &  $24.86^{+0.04}_{-0.04}$ \\[0.10cm]
		Q1422-md145a &  14:24:35.52 &  +22:57:18.53 &           ... &  $27.78^{+0.10}_{-0.09}$ &  $26.75^{+0.03}_{-0.03}$ &  $26.76^{+0.04}_{-0.04}$ \\[0.10cm]
		Q1422-md145b &  14:24:35.56 &  +22:57:20.26 &         2.800 &  $25.76^{+0.05}_{-0.05}$ &  $25.78^{+0.04}_{-0.04}$ &  $25.34^{+0.03}_{-0.03}$ \\[0.10cm]
		Q1422-md145c &  14:24:35.53 &  +22:57:19.36 &         2.800 &  $27.92^{+0.12}_{-0.11}$ &  $27.45^{+0.06}_{-0.06}$ &  $27.69^{+0.10}_{-0.09}$ \\[0.10cm]
		Q1549-C10a &  15:51:48.45 &  +19:09:24.67 &         3.189 &  $26.84^{+0.07}_{-0.07}$ &  $27.37^{+0.13}_{-0.12}$ &  $27.64^{+0.23}_{-0.19}$ \\[0.10cm]
		Q1549-C10b &  15:51:48.42 &  +19:09:25.01 &         3.189 &  $24.73^{+0.03}_{-0.03}$ &  $24.90^{+0.03}_{-0.03}$ &  $24.72^{+0.04}_{-0.04}$ \\[0.10cm]
		Q1549-C27 &  15:52:07.05 &  +19:12:19.31 &         2.926 &  $24.42^{+0.02}_{-0.02}$ &  $24.62^{+0.03}_{-0.03}$ &  $24.26^{+0.03}_{-0.03}$ \\[0.10cm]
		Q1549-C8 &  15:51:45.39 &  +19:08:49.85 &         2.937 &  $24.99^{+0.03}_{-0.03}$ &  $25.15^{+0.05}_{-0.05}$ &  $24.64^{+0.04}_{-0.04}$ \\[0.10cm]
		Q1549-D7 &  15:51:46.25 &  +19:09:50.10 &         2.936 &  $24.18^{+0.02}_{-0.01}$ &  $24.10^{+0.01}_{-0.01}$ &  $23.73^{+0.01}_{-0.01}$ \\[0.10cm]
		Westphal-CC32a &  14:18:14.48 &  +52:28:07.26 &         3.192 &  $26.01^{+0.03}_{-0.03}$ &  $25.92^{+0.03}_{-0.03}$ &  $25.71^{+0.04}_{-0.03}$ \\[0.10cm]
		Westphal-CC32b &  14:18:14.46 &  +52:28:06.99 &         3.192 &  $24.68^{+0.02}_{-0.02}$ &  $24.67^{+0.02}_{-0.02}$ &  $24.49^{+0.03}_{-0.03}$ \\[0.10cm]
		Westphal-CC41a &  14:18:20.55 &  +52:29:21.14 &         3.027 &  $26.01^{+0.05}_{-0.05}$ &  $26.00^{+0.06}_{-0.06}$ &  $25.62^{+0.05}_{-0.05}$ \\[0.10cm]
		Westphal-CC41b &  14:18:20.49 &  +52:29:21.48 &         3.027 &  $26.00^{+0.05}_{-0.05}$ &  $25.84^{+0.05}_{-0.05}$ &  $25.57^{+0.05}_{-0.05}$ \\[0.10cm]
		Westphal-CC46 &  14:18:00.20 &  +52:29:53.02 &         3.261 &  $23.96^{+0.01}_{-0.01}$ &  $23.76^{+0.01}_{-0.01}$ &  $23.58^{+0.01}_{-0.01}$ \\[0.10cm]
		Westphal-CC53 &  14:18:22.15 &  +52:30:19.70 &         2.807 &  $25.22^{+0.04}_{-0.04}$ &  $25.08^{+0.04}_{-0.04}$ &  $24.48^{+0.03}_{-0.03}$ \\[0.10cm]
		Westphal-DD28 &  14:18:24.87 &  +52:29:27.32 &         3.021 &  $24.49^{+0.02}_{-0.02}$ &  $24.22^{+0.02}_{-0.02}$ &  $23.83^{+0.01}_{-0.01}$ \\[0.10cm]
		Westphal-MM38 &  14:18:04.07 &  +52:29:54.94 &         2.925 &  $25.53^{+0.03}_{-0.03}$ &  $25.16^{+0.03}_{-0.03}$ &  $24.77^{+0.03}_{-0.03}$ \\[0.10cm]
		\bottomrule
	\end{tabular}
	\begin{flushleft}
		$^a$ {The field the object is located in, the object name, and a letter corresponding to the subcomponent in Figure \ref{fig:nondets}. A subcomponent label is omitted in the case of single-component morphology.}
		$^b$ {Systemic redshift from \citetalias{Steidel2018}.}
		$^c$ {Systemic redshifts are omitted for components predicted as low redshift based on their $V_{606}J_{125}H_{160}$ colors.}
	\end{flushleft}
	\label{table:nondet_phot}
\end{table*}

\end{CJK*}
\end{document}